\RequirePackage{fix-cm}

\documentclass[twocolumn,epjc3]{svjour3}
\usepackage[english]{babel}
\usepackage[T1]{fontenc}
\usepackage[utf8]{inputenc}
\RequirePackage{graphicx}
\usepackage{amsmath}
\usepackage{amssymb}
\usepackage{booktabs}
\RequirePackage[numbers,compress]{natbib}

\usepackage{hyperref}
\hypersetup{
    colorlinks=true,
    linkcolor=blue,
    filecolor=magenta,
    urlcolor=cyan,
    citecolor=cyan
}

\journalname{Eur. Phys. J. C}

\def\be{\begin{equation}}
\def\ee{\end{equation}}

\def\bea{\begin{eqnarray}}
\def\eea{\end{eqnarray}}

\begin{document}

\title{Logarithmic corrections to Newtonian gravity and Large Scale Structure}

\author{Salvatore Capozziello\thanksref{e1,addr1,addr2,addr3,addr4}
        \and
        Mir Faizal\thanksref{e2,addr5,addr6,addr7}
        \and
        Mir Hameeda\thanksref{e3,addr8,addr9}
        \and
        Behnam Pourhassan\thanksref{e4,addr10}
        \and
        Vincenzo Salzano\thanksref{e5,addr11} 
}

\thankstext{e1}{e-mail: capozziello@unina.it}
\thankstext{e2}{e-mail: mirfaizalmir@googlemail.com}
\thankstext{e3}{e-mail: hme123eda@gmail.com}
\thankstext{e4}{e-mail: b.pourhassan@du.ac.ir}
\thankstext{e5}{e-mail: vincenzo.salzano@usz.edu.pl}

\institute{Department of Physics ``E. Pancini'', University of Naples ``Federico II'', Naples, Italy \label{addr1}
\and
INFN Sez. di Napoli, Compl. Univ. di Monte S. Angelo, Edificio G, Via	Cinthia, I-80126, Naples, Italy \label{addr2}
\and
Scuola Superiore Meridionale, Largo San Marcellino 10, I-80138, Naples, Italy \label{addr3}
\and
Laboratory for Theoretical Cosmology, Tomsk State University of Control Systems and Radioelectronics (TUSUR), 634050 Tomsk, Russia \label{addr4}
\and
Irving K. Barber School of Arts and Sciences, University of British Columbia - Okanagan, 3333 University Way, Kelowna, British Columbia V1V 1V7, Canada \label{addr5}
\and
Department of Physics and Astronomy, University of Lethbridge, Lethbridge, Alberta, T1K 3M4, Canada \label{addr6}
\and
Canadian Quantum Research Center 204-3002 32 Ave Vernon, BC V1T 2L7 Canada \label{addr7}
\and
Department of Physics, Government Degree College Tangmarg, Kashmir 193402, India \label{addr8}
\and
Visiting Associate, IUCCA,  Pune,  41100, India \label{addr9}
\and
School of Physics, Damghan University, Damghan, 3671641167, Iran \label{addr10}
\and
Institute of Physics, University of Szczecin, Wielkopolska 15, 70-451 Szczecin, Poland \label{addr11}
}

\date{Received: date / Accepted: date}

\maketitle

\begin{abstract}
Effects from nonstandard corrections to Newtonian gravity, at large scale, can be investigated using the  cosmological  structure formation. In particular,  it is possible to show if and how a logarithmic correction (as that induced from nonlocal gravity) modifies the clustering properties of galaxies and of clusters of galaxies. The thermodynamics of such systems can be used to obtain important information about the effects of such modification on clustering. We will compare its effects with observational data and it will be demonstrated  that the observations seem to point to a characteristic scale where such a logarithmic correction might be in play at galactic scales. However, at larger scales such statistical inferences are much weaker, so that a fully reliable statistical evidence for this kind of corrections cannot be stated without further investigations and the use of more varied and precise cosmological and astrophysical probes.
\end{abstract}

\section{Introduction}

The  clustering of galaxies is the main mechanism to address the large scale structure of the Universe.  Such clustering mechanism can be  studied confronting numerical  simulations  \citep{a1,a2} with observations \citep{a3,a4}, using the local matter distributions of galaxies organized in  groups, filaments and clusters. However, to analyze numerically the large scale structure formation, it is possible to approximate galaxies as point-like particles, and then study the clustering adopting the standard formalism of statistical mechanics. This approximation is valid because the distance between galaxies is  much larger than their proper size. In fact, it has been demonstrated that the clustering mechanism can be studied considering a  quasi-equilibrium description because the macroscopic quantities  change very slowly as compared with the  local relaxation time scales  \citep{cos1,cos2,cos3,cos4}. Pressure, average density and  average temperature of clusters  are the macroscopic quantities to be taken into account. Here the temperature is the effective one obtained by the kinetic theory of gases, for a gas of galaxies (each galaxy being approximated as a point particle). So, by this quasi-equilibrium description, the  clustering of galaxies  can be  dealt as a thermodynamic system \citep{cosa1,sas2}. Specifically, one has to  adopt a gravitational partition function \citep{sas,sas2} where  point-like galaxies  gravitationally interact. Such a  partition function can  be also used to study phase transitions for systems of interacting galaxies \citep{sin12,3,2b}.

It has to be pointed out   that such a   gravitational partition function diverges because  the extended structure of galaxies has been neglected in this approximation.  However, these divergences can be removed by adding a  softening parameter, which accounts for the extended structure of  galaxies as reported in  \citep{ahm10}. In a recent study it has also been proved that the nonlocal gravity is divergence free \cite{reference_epjc}. Thermodynamics for  systems of galaxies is obtained from the partition function (regularized with a softening parameter), and this can in turn be used to study the clustering properties \citep{sas84}.
It has been observed that any modification of  gravitational partition function by  a softening parameter, in turn, modifies the thermodynamic quantities of a given  system. Vice versa, because these thermodynamic quantities are related to the clustering parameter, it also changes the clustering parameter itself.
In fact, it can be demonstrated that the modification to the   skewness and kurtosis of clustering systems  occur due to such a softening parameter   \citep{ahm02}.  This formalism has also been used to demonstrated that galaxy clusters are surrounded by  halos, and this has been done adopting a wide range of samples  \citep{10,20}. Furthermore the isothermal compressibility  gives also  information on the clustering of galaxies \citep{ahm06}.

Specifically,  the clustering of galaxies occurs due to the  gravitational force between them. However, it is now well-known by a wide range of cosmological observations (e.g. Type Ia Supernovae, Baryon Acoustic Oscillations and Cosmic Microwave Background) that the Universe is undergoing an accelerating  expansion \citep{1ab,6ab,Aghanim:2018eyx}. This cosmic expansion is pulling the galaxies away from each other. So, the cosmic expansion seems to oppose the gravitational force, and thus it is important to include its effect in the gravitational partition function. It has been argued that the cosmological constant related to the accelerated expansion can produce important local  effects   on the  scale of galaxy clusters  \citep{cosd1,cosd2,cosd3,cosd4,cosd5}. These effects can be investigated using model of interacting  dark matter and dark energy \citep{cosd1,cosd2}. It is also possible to   constrain the dark energy from the clustering of    galaxies \citep{cosd51}. The   cosmic expansion  can also change the dynamics at the scale of  galaxy clusters \citep{cosd6}.  Such a modification of cluster dynamics  has also been investigated using   different   groups of galaxies  \citep{cosd71}.  Finally, one can state that the cosmological constant  has   consequences for  the  formation of galaxy  clusters \citep{cosd7}. In summary,  it is important to study the effects of  cosmic expansion at the scale of clusters. From a more technical point of view, it has to be noted that the Fisher matrix formalism has been used to analyze the effects of cosmic expansion on clusters \citep{cosd8}. 

Thus, it is possible  to obtain information about the cosmological constant, and, in general, about dark energy dynamics, from the local clustering of galaxies. So, it is important to incorporate the effects of cosmic expansion in the statistical mechanical description of the clustering of galaxies. In fact, an accurate measurement of the statistics of galaxies can be used to constraint the value of the cosmological constant \citep{cosd9}, and the distribution function of galaxies can be  used to constraint the amount of the  dark energy \citep{cosd91}.

As such a statistical distribution function can be obtained from the gravitational partition function \citep{cosa1,sas2},   it is important to incorporate the effects of cosmic expansion in the  partition function. So,  the  modification of the gravitational partition function by the cosmological constant term  has also to be considered, and this modified gravitational partition function can  be used to study the  clustering process in an expanding Universe \citep{1b}.
This can be technically   done by first analyzing the Helmholtz free energy and entropy for a system of galaxies. Then the clustering parameter for this system can  be obtained, and it has been observed that this clustering parameter depends on the value of the cosmological constant. The modification of the gravitational partition function with a time dependent cosmological constant can be  also  used to study the effects of dynamical dark energy on clustering of galaxies \citep{1}. It has been observed that the  correlation function for this system are consistent with observations.

However, it is possible to obtain cosmic expansion without considering dark energy related to some new fundamental particle. In fact,    large scale nonlocality  related to some effective gravitational potential seems a realistic effect of gravitational field at infrared scales as recently reported in  several papers, see for example \citep{50acd,50ab,Bahamonde,Nunes,Bajardi}. It has been argued that such  nonlocalities can occur  due to some quantum gravitational effects  \citep{50acd,50ab,Modesto}. They  can be relevant  also in gravitational waves modes and polarizations \citep{Capriolo,50ac}. It has been demonstrated that  string field theory produces nonlocal actions for the component fields, and these nonlocal actions can also be used to construct  nonlocal cosmological models \citep{sftn1,sftn2}. It is also possible to construct models of nonlocal cosmology from loop quantum gravity \citep{lqgn1,lqgn2}.  This can be done adopting a sort of condensation of states in loop quantum gravity. In summary, it is possible to  obtain nonlocal cosmologies using  quantum gravitational effects at ultraviolet scales that propagate up to infrared scales.

In other words, as nonlocal cosmology can explain the cosmic  expansion without dark energy \citep{50acd,50ab,Bahamonde,Nunes,Bajardi}, such nonlocal corrections  can also  be used to incorporate the effects of cosmic expansion in the gravitational partition function. This one can then be used to analyze the effects of nonlocality on clustering of galaxies. It may be noted that even though general relativity is  strongly   constrained at the  Solar System scales by several experiments and observations \citep{50,50a,50b}, nonlocal  modification to general relativity can  occur at large scales starting from galactic one \citep{Kostas}. At the same time, nonlocal gravity is safe from possible issues related to oscillating features on small scales, falling in the regime where such oscillations are subdominant \cite{Lazar:2020gsx}.
It has also been proposed that such nonlocality could be constrained from the clustering of galaxies \citep{2019JCAP}. Thus, it is important to analyze the effect of nonlocal modification of gravity on  the gravitational partition function and point out how it affects the clustering process. Such a modification can be analyzed using  nonlocal corrections to the gravitational potential \citep{Kostas,40a,40b,40c,40d}. In general, the  nonlocal gravitational partition function can  be used to analyze the effects of nonlocality on clustering of galaxies. Although our main motivation for such a study is related to testing nonlocal gravity, we must note here that the logarithmic correction induced by such a theory is shared also by other alternative models of gravity, e.g. MOND \citep{Shapiro:2004ch,Famaey:2011kh}, Randall-Sundrum brane scenario \citep{Park:2002rb}, or other string-theoretical motivations \citep{Soleng:1993yr} which have been also tested with internal dynamics of spiral galaxies \citep{Fabris:2007df}. It may be noted that there is an intrinsic non-locality in string theory \citep{Calcagni:2013eua,Biswas:2012ka}. Furthermore, even MOND has been related to nonlocality \citep{Soussa:2003vv,Deffayet:2011sk, Deffayet:2014lba,Tan:2018bfp,Kim:2016nnd}. This could be due to the fact that nonlocality \citep{50acd,50ab}, and several of these modifications, are motivated from quantum corrections to the original Einstein-Hilbert action. Even MOND can be obtained as a quantum correction in the  Verlinde formalism \citep{Bagchi:2017jfl}. Thus, we expect some kind of universal form of correction to occur in most of them. In this perspective, our results will be more general and not strictly related to one particular scenario.

In Sec.~2, we discuss  the  clustering mechanism. After defining the nonlocal gravitational partition function, we consider the related thermodynamics and spatial correlation function. Observational data analysis is described in detail in Sec.~3. Specifically, we take into account data from galaxies and clusters and study the effects of nonlocality on the clustering mechanism. Results are discussed in Sec.~4, while conclusions are drawn in Sec.~5.

\section{Modeling Clusters of galaxies}

In general, a nonlocal modification of the 
Hilbert-Einstein action can have the following form \citep{50acd,50ab}
\begin{equation}\label{actionDeser}
\mathcal{S} =\frac{1}{2 \kappa ^2} \int d^4 x \sqrt{-g} \left[ R 
\left( 1 + f(\square ^{-1} R) \right) \right]  \,,
\end{equation}
where $R$ is the Ricci scalar and $f(\square^{-1}R)$ is an arbitrary 
function, called \textit{distortion function}, of the nonlocal term 
$\square ^{-1}R$, which is explicitly given by the retarder Green's 
function
\begin{equation}\label{Greenfunc}
\mathcal{G}[f](x)=(\square ^{-1}f)(x) = \int d^4x' 
\sqrt{-g(x')}f(x')G(x,x')\,.
\end{equation}
Setting $f(\square^{-1}R)=0$, the above action is equivalent to 
the Einstein-Hilbert one.  The nonlocality is introduced by the inverse of the d'Alembert operator. In order to "localize" the action, an auxiliary scalar field can be introduced so that $\square^{-1}R=\phi$ and then formally $\square \phi=R$. As a consequence, characteristic lengths related to the nonlocal terms naturally come out \citep{Kostas}.

The weak-field limit of such a dynamics gives rise to corrections to the Newtonian potential which are interesting at large scales. These corrections can be polynomial or logarithmic and, in general, introduce characteristic scales. For a detailed discussion on this point see \citep{Kostas}.

In the framework of this theory, let us  now model the clustering of galaxies using the gravitational partition function which we are going to define. We will explicitly adopt the nonlocal gravity, as discussed in \citep{40b,40c,40d},  to obtain the corrections to such a gravitational partition function. After  analyzing  the effects of nonlocality, the nonlocal gravitational partition function will be used to analyze the  clustering of galaxies.  This can be done by considering  the thermodynamics of this system, and then relating it to the clustering parameter. Then we will calculate the spatial correlation function.

\subsection{Nonlocal Gravitational Partition Function}

Let us take into account   a system with a large number of galaxies. This system can be analyzed using  a quasi-equilibrium description, where    the change in  macroscopic quantities is slower than  the  local relaxation time scales  \citep{cos1,cos2,cos3,cos4}. It is possible analyze such a system adopting    an ensemble of cells, with  same volume $V$, or radius $R_1$ and average density$\rho$. As the    number of galaxies and their total energy can vary between  these cells, the system can be analyzed in the framework of the grand canonical ensemble. Thus, it is possible to define a gravitational partition function.   In this picture,    a system of $N$ galaxies  of mass $m$ interacting with  a nonlocal   gravitational potential  can be written as \citep{sas,sas2}
\begin{eqnarray}
Z(T,V)&=&\frac{1}{\lambda^{3N}N!}\int d^{3N}pd^{3N}r  \nonumber \\
&\times& \exp\biggl(-\biggl[\sum_{i=1}^{N}\frac{p_{i}^2}{2M}+\Phi_{nl}(r)\biggr]T^{-1}\biggr),
\end{eqnarray}
where $p_{i}$ are the momenta of different galaxies  and $\Phi_{nl}$ is the nonlocal gravitational potential energy. Here $T$ is the average temperature, which is obtained from the kinetic theory of gases where each particle is represented by a galaxy.  Now
integrating on the   momentum space, we can write this gravitational partition function as
\begin{eqnarray}
Z_N(T,V)=\frac{1}{N!}\left(\frac{2\pi mT}{\Lambda^2}\right)^{3N/2}Q_N(T,V),
\end{eqnarray}
where $Q_{N}(T,V)$,  the configuration  integral, can be expressed as
\begin{equation}
Q_{N}(T,V)=\int....\int \prod_{1\le i<j\le N} \exp[-\frac{\Phi_{nl}(r_{ij})}{T}]d^{3N}r. \label{q1}
\end{equation}
The nonlocal gravitational potential energy $\Phi_{nl}(r_{1}, \dots, r_{N})$ is
a function of the relative position vector $r_{ij}=|r_{i}-r_{j}|$, as both the local and nonlocal terms represent central forces. The total potential energy  can be obtained by summing up  the  potential energies. So, we can write the total nonlocal  potential energy $\Phi_{nl}(r_{1}, r_{2}, \dots, r_{N})$  as
\begin{align}
&\Phi_{nl}(r_{1}, \dots, r_{N})=\sum_{1\le i<j\le N}\Phi_{nl}(r_{ij})=\sum_{1\le i<j\le N}\Phi_{nl}(r).
\end{align}
 We can write such  a nonlocal gravitational potential energy for galaxies as \citep{40b,40c,40d}
\begin{eqnarray}\label{eq:nonlocal_pot}
(\Phi_{i,j})_{nl}=-\frac{Gm^2}{r_{ij}}+\frac{Gm^2}{\lambda}ln\left(\frac{r_{ij}}{\lambda}\right)
\end{eqnarray}
Since galaxies have an  extended structure, we need to modify the gravitational potential energy by a  softening parameter $\epsilon$ \citep{ahm10,ahm06}. Thus, we can write 
\begin{eqnarray}
(\Phi_{i,j})_{nl}=-\frac{Gm^2}{(r_{ij}^2+\epsilon^2)^{1/2}}+\frac{Gm^2}{\lambda}ln\left(\frac{(r_{ij}^2+\epsilon^2)^{1/2}}{\lambda}\right)
\end{eqnarray}

Now  we can use  a nonlocal   two-point interaction function for galaxies $f_{ij}=e^{-(\Phi_{ij})_{nl}/T}-1$ to analyze this system. It vanishes  in absence of interactions,  and it is non-zero only for interacting galaxies. Thus,  we can now express the nonlocal  configuration integral as
\begin{eqnarray}
Q_{N}(T,V)&=&\int....\int \biggl[(1+f_{12})(1+f_{13})(1+f_{23})\nonumber\\
&& \dots (1+f_{N-1,N})
\biggr]d^{3}r_{1}d^{3}r_{2}\dots d^{3}r_{N},\label{q2}
\end{eqnarray}
where we can express the nonlocal two-point  interaction Mayer function as
\begin{align}
&f_{ij}=\left(\frac{(r_{ij}^2+\epsilon^2)^{1/2}}{\lambda}\right)^{\frac{Gm^2}{T\lambda}}\exp\left(\frac{Gm^2}{T(r_{ij}^2+\epsilon^2)^{1/2}}\right)-1
\end{align}
It is worth noticing that such an expression for clustering integral  has been discussed for local gravitational interactions \citep{ahm02,ahm06}. Here we have obtained the contributions to such Mayer function from nonlocal gravitational interactions \citep{40b,40c,40d}.  We can expand this nonlocal two-point interaction   Mayer function  as a power series. So, we can write $f_{ij}$ as
\begin{align}
f_{ij}&=-1+\left(\frac{(r_{ij}^2+\epsilon^2)^{1/2}}{\lambda}\right)^{\frac{Gm^2}{T\lambda}}
\nonumber\\
&\cdot\sum_{n=0}^{\infty}\frac{1}{n!}\left(\frac{Gm^2}{T}\right)^n\frac{1}{(r_{ij}^2+\epsilon^2)^{n/2}}
\end{align}
Now using  $l={Gm^2}/({T\lambda})$, we can write
$Q_{2}(T,V)$ as
\begin{eqnarray}
Q_{2}(T,V)&=&V^2\sum_{n=0}^{\infty}\frac{1}{n!}\bigl(\frac{3Gm^2}{2R_{1}T}\bigr)^n\bigl(\frac{\epsilon}{\lambda}\bigr)^l\bigl(\frac{2R_{1}}{3\epsilon}\bigr)^n \nonumber \\
&\cdots& 2F1\left(\frac{3}{2},\frac{n+l}{2};\frac{5}{2};-\left(\frac{R}{\epsilon}\right)^2\right)
\nonumber \\ &=& V^2\big(\sum_{n=0}^{\infty}\alpha_{nl}x^{n}\big),
\end{eqnarray}
where $\alpha_{nl}$ is given by
\begin{eqnarray}
\alpha_{nl}&=& \frac{1}{n!}\biggl(\frac{\epsilon}{\lambda}\biggr)^l\biggl(\frac{2R_{1}}{3\epsilon}\biggr)^{n} F \biggl(\frac{3}{2},\frac{n+l}{2};\frac{5}{2};-\frac{R_{1}^2}{\epsilon^2}\biggr).
\end{eqnarray}
It may be noted here  that $ F ({3}/{2},{n+l}/{2};{5}/{2};-{R_{1}^2}/{\epsilon^2})$ is the hypergeometric function.
Now using $R_{1}\sim \rho^{-1/3}$, we observe that
${3Gm^{2}}/{2R_{1}T}={3Gm^{2}}/{2\rho^{-1/3}T}={3}G\\m^{2}\rho^{1/3}T^{-1}/2.
$ So,  using the scale invariance,  $\rho\to\lambda^{-3}\rho$, $T\to\lambda^{-1}T$ and $r\to\lambda r$, we  can define $x$ as
\begin{eqnarray}
x = \frac{3}{2}(Gm^2)^{3}\rho T^{-3}=\beta\rho T^{-3},
\end{eqnarray}
where $\beta= {3}(Gm^{2})^{3}/2 $. From this expression for $x$, we can obtain any   general configuration integral. 
Using the so called {\it dilute approximation},   we  can assume ${(\Phi_{i,j})_{nl}}/{T}$   very small and  take only the first term of the exponential expansion,
\begin{eqnarray}
f_{ij}=\frac{Gm^2}{T\sqrt{r_{ij}^2+\epsilon^2}}-\frac{Gm^2}{T\lambda}\ln{\sqrt{r_{ij}^2+\epsilon^2}}+\frac{Gm^2}{T\lambda}\ln{\lambda}
\end{eqnarray}
Thus, calculating the general configuration integral, we obtain
\begin{eqnarray}
Q_{N}(T,V)=V^N\left(1+\alpha x\right)^{N-1}
\end{eqnarray}
where
\begin{eqnarray}\label{eq:alpha}
\alpha&=&\frac{2R}{3\lambda}\left(
\ln \frac{\lambda}{R}+
\frac{3\lambda}{2R}\sqrt{\frac{\epsilon^2}{R^2}+1}+
\frac{\epsilon^3}{R^3}\tan^{-1} \frac{R}{\epsilon}\right. \\
&-&\left.\frac{3\lambda\epsilon^2}{2R^3} \ln\frac{\sqrt{\frac{\epsilon^2}{R^2}+1}+1}{\frac{\epsilon}{R}}-
\frac{1}{2}\ln\left(\frac{\epsilon^2}{R^2}+1\right)-\frac{\epsilon^2}{R^2}+\frac{1}{3}\right) \nonumber
\end{eqnarray}
Now the gravitational partition function for this dilute gravitating system can be written as
\begin{eqnarray}
Z_N(T,V)=\frac{1}{N!}\big(\frac{2\pi m T}{\Lambda^2}\big)^{3N/2}V^{N}\big(1+\alpha x\big)^{N-1}
\end{eqnarray}
This expression for the gravitational partition function can be used to analyze the effects of nonlocality on the clustering of galaxies.

\subsection{Thermodynamics}
It has to be noted that the gravitational partition function has already been used to study the thermodynamics of  systems of galaxies,   see \citep{cosa1} and  \citep{sas2}.
We can thus adopt the nonlocal gravitational  partition function to analyze the thermodynamics of this nonlocal system. The
Helmholtz free energy $F=-T\ln Z_{N}(T,V)$ is
\begin{eqnarray}
F=  -T\ln \biggl(\frac{1}{N!}\big(\frac{2\pi mT}{\Lambda^2}\big)^{3N/2}V^N\big(1+\alpha x\big)^{N-1}\biggr)\,.
\end{eqnarray}

\begin{figure*}
\begin{center}
\includegraphics[width=0.9\textwidth]{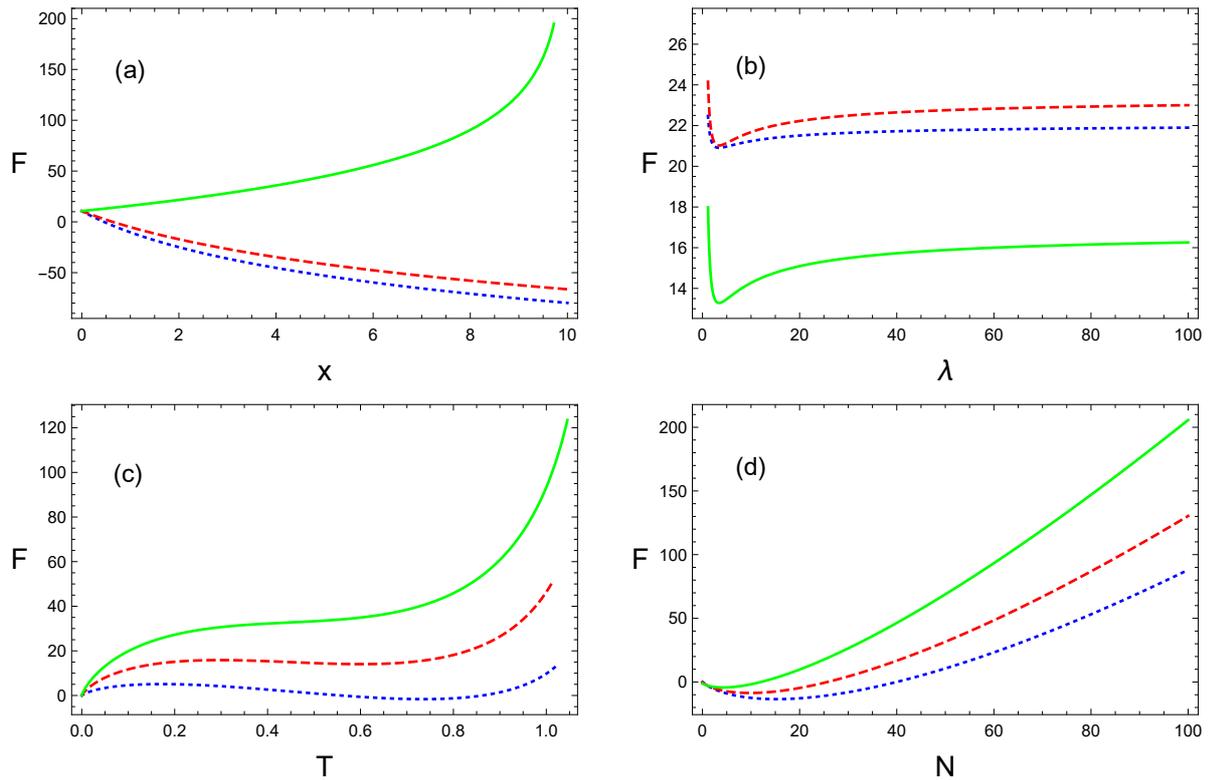}
\end{center}
\caption{Helmholtz free energy dependence on: 
(a) $x$ with: $T=1$, $N=50$, $\lambda=2/3$ (solid green), $\lambda=1$ (dashed red), $\lambda=\infty$ (dotted blue); 
(b) $\lambda$ with: $x=1$, $N=50$, $T=0.6$ (solid green), $T=0.4$ (dashed red), $T=0.2$ (dotted blue);
(c) $T$ with: $x=1$, $N=60$ (solid green), $N=40$ (dashed red), $N=20$ (dotted blue); 
(d) $N$ with: $T=1$, $\lambda=0.5$; $x=1$ (solid green), $x=0.5$ (dashed red), $x=0$ (dotted blue). We have  assumed unit values for all the other parameters.}
 \label{fig1}
\end{figure*}

In the plots of  Fig. \ref{fig1},  we analyze the  dependence of  the Helmholtz free energy on various parameters. The  dependence of Helmholtz free energy on    $x$ is represented in  Fig. \ref{fig1} (a). We can observe that the  Helmholtz free energy is a  decreasing function of $x$ for large value of the nonlocality parameter $\lambda$ and  an increasing  function of $x$ for smaller value of the nonlocality parameter $\lambda$. So, the value of nonlocality parameter changes the dependence of the Helmholtz free energy on $x$.  In  Fig. \ref{fig1} (b), the  Helmholtz free energy is  plotted  in terms of $\lambda$. We can see how there is a minimum for the 
Helmholtz free energy, and this minimum seems to occur at a specific value of $\lambda$. This minimum does not seem to change with the  temperatures of the system. This observation can explain the change in the behavior of the Helmholtz free energy with the change in the value of $\lambda$, as was observed in  Fig. \ref{fig1} (a). 
In  Fig. \ref{fig1} (c), we plot the dependence of the Helmholtz free energy on the  temperature of the system. Here the temperature corresponds to the temperature of a gas of galaxies, with each galaxy acting as a particle analog. We observe that generally the  Helmholtz free energy increases with the temperature, but the rate of this increase depends on the number of galaxies. As we have plotted the dependence on temperature by considering the system as a gas of galaxies, we  also analyze the dependence of the Helmholtz free energy on the number of galaxies in   Fig. \ref{fig1} (d). We observe that  the Helmholtz free energy first decreases for a relatively small number of galaxies, becoming negative, and then increases as the number of galaxies becomes more than a certain value. 

The entropy $S$ can now be calculated from this Helmholtz free energy, as
\begin{align}
S&= -\biggl(\frac{\partial F}{\partial T}\biggr)_{N,V}
 \nonumber \\
 &= N\ln (\rho^{-1}T^{3/2})+(N-1)\ln \big(1+\alpha x\big) \nonumber \\ 
 &-3N
 \frac{\alpha x}{1+\alpha x}
 +\frac{5}{2}N+\frac{3}{2}N\ln \big(\frac{2\pi m}{\Lambda^2}\big).
\end{align}
Now,  for large $N$, using $N-1\approx N$, we obtain
\begin{eqnarray}
\frac{S}{N}=\ln (\rho^{-1}T^{3/2})+\ln \big(1+\alpha x\big)-3B_l+\frac{S_{0}}{N},
\end{eqnarray}
where $S_{0}=\frac{5}{2}N+\frac{3}{2}N\ln \big(\frac{2\pi m}{\Lambda^2}\big)$.
where
\begin{eqnarray}
B_l=\frac{\alpha x}{1+\alpha x}.
\end{eqnarray}
This is the general clustering parameter for a system of galaxies interacting in presence of  nonlocal  gravity.
The internal energy $U =  F+TS$   of a system of galaxies can now be written  as
\begin{eqnarray}
U = \frac{3}{2}NT\big(1-2B_l\big).
\end{eqnarray}

\begin{figure*}
 \begin{center}
\includegraphics[width=0.9\textwidth]{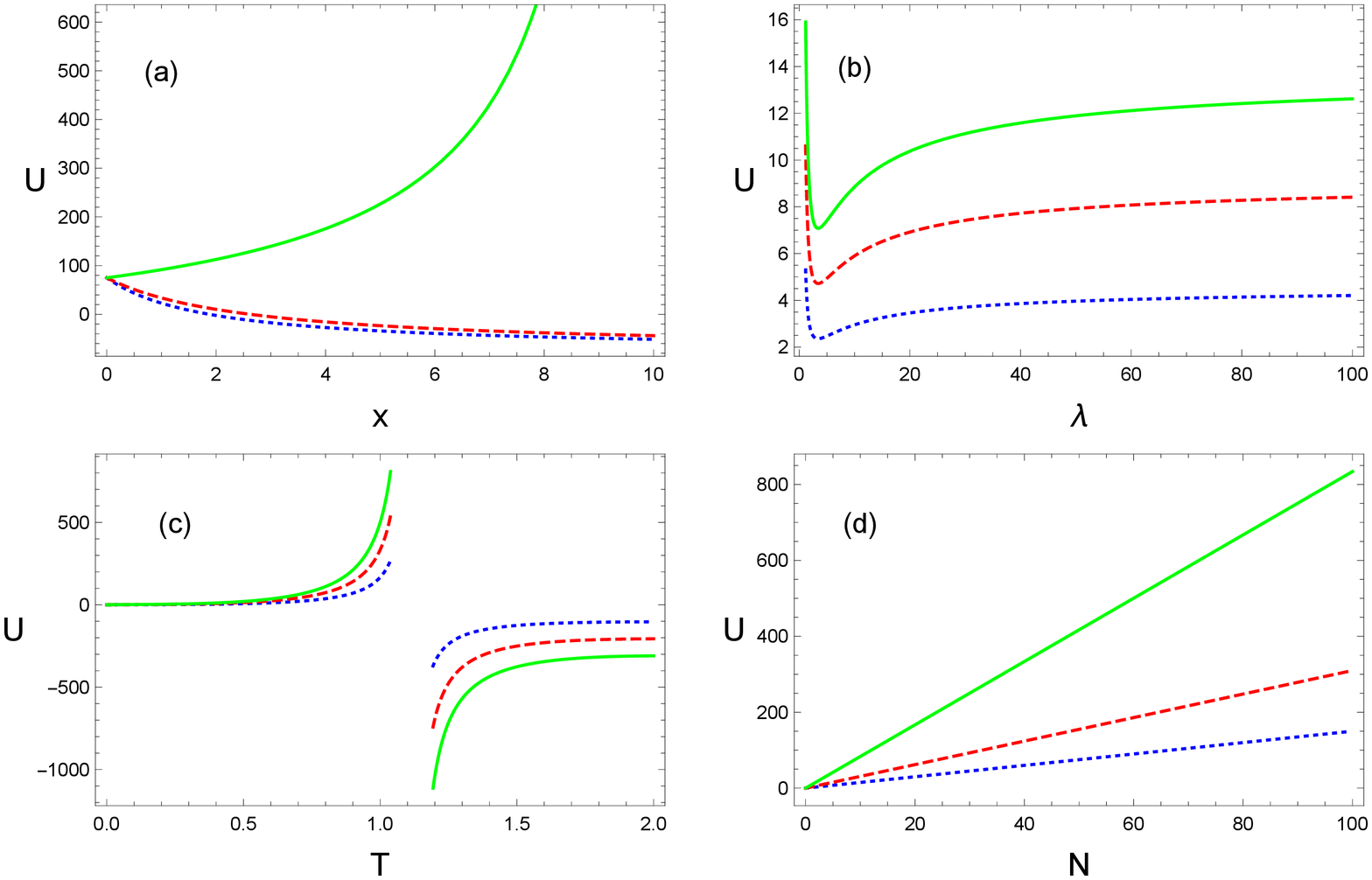}
 \end{center}
\caption{Internal energy dependence on: 
(a) $x$ with: $T=1$, $N=50$, $\lambda=2/3$ (solid green), $\lambda=1$ (dashed red), $\lambda=0$ (dotted blue); 
(b) $\lambda$ with: $x=1$, $N=50$, $T=0.6$ (solid green), $T=0.4$ (dashed red), $T=0.2$ (dotted blue);
(c) $T$ with: $x=1$, $N=60$ (solid green), $N=40$ (dashed red), $N=20$ (dotted blue); 
(d) $N$ with: $T=1$, $\lambda=0.5$; $x=1$ (solid green), $x=0.5$ (dashed red), $x=0$ (dotted blue). We have  assumed unit values for all the other parameters.}
 \label{fig2}
\end{figure*}

We have  plotted the internal energy of a system of galaxies in    Fig. \ref{fig2}. Its   behavior is  analyzed by investigating its dependence on various  parameters.  In Fig. \ref{fig2} (a), we investigate how the internal energy of the system varies with $x$. We observe that it is an  increasing function of $x$ for smaller values of $\lambda$, and a decreasing function of $x$ for larger values of $\lambda$.  In  Fig. \ref{fig2} (b), we analyze the dependence of the internal energy on the    nonlocality  parameter $\lambda$. We observe that there is a   minimum for internal energy, and again this minimum does not seem to depend on the temperature of the system.  We  also investigate the dependence of the internal energy on the temperate of the system in Fig. \ref{fig2} (c). It is observed that  there is a discontinuity in the dependence  of   internal energy  on temperature, which occurs independently of the number of galaxies. In Fig. \ref{fig2} (d), the dependence of internal energy on the number of galaxies is plotted. 
It is observed that internal energy increases with the increase in the number of galaxies. However, the  rate of increase is larger for larger values of $x$.

\begin{figure*}
 \begin{center}
\includegraphics[width=0.9\textwidth]{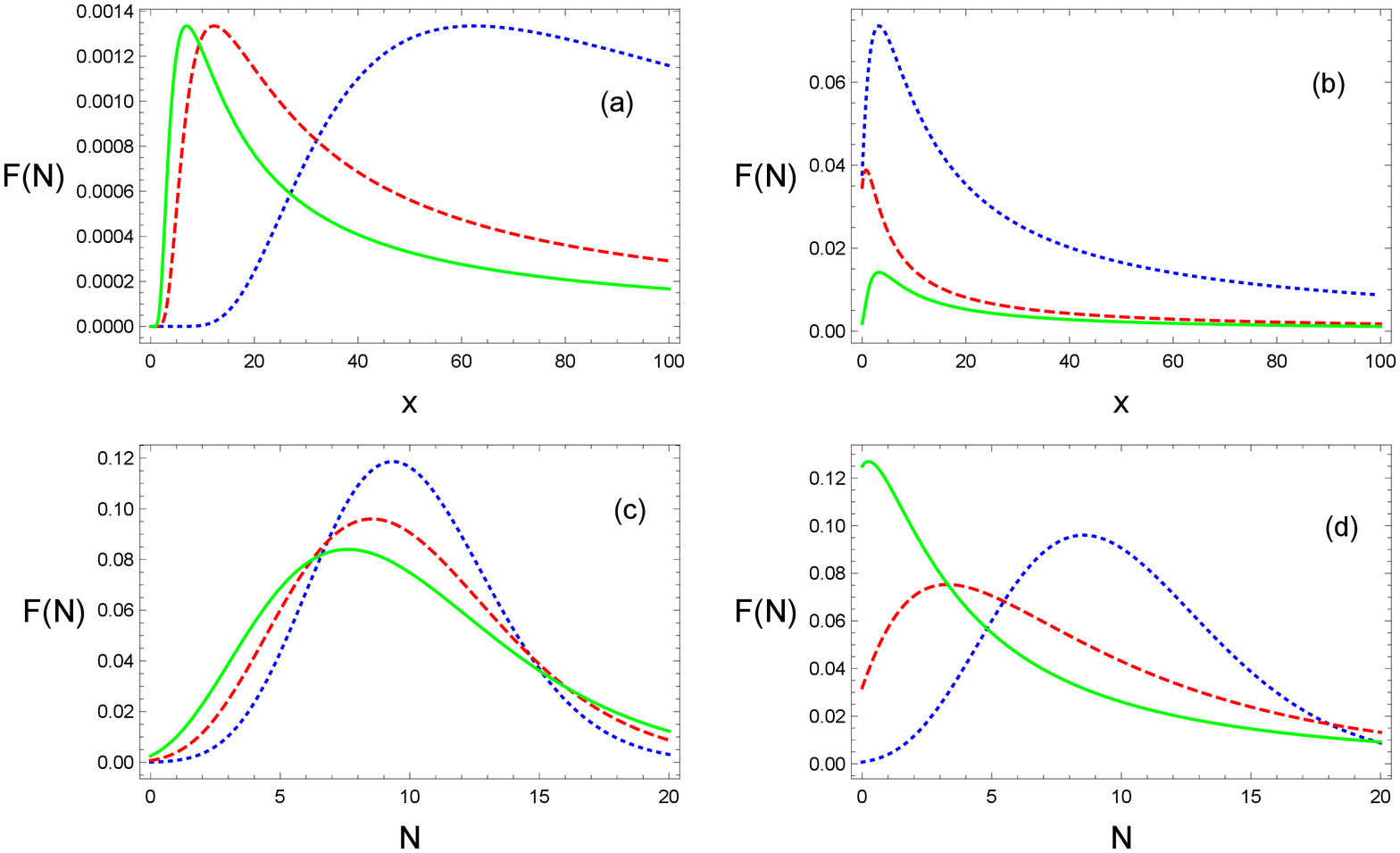}
 \end{center}
\caption{The distribution function dependence on: (a) $x$ with: $N=50$, $\bar{N}$=10, $\lambda=10$ (solid green), $\lambda=1$ (dashed red), $\lambda=0.75$ (dotted blue); 
(b) $x$ with: $\bar{N}$=10, $\lambda=1$, $N=20$ (solid green), $N=10$ (dashed red), $N=5$ (dotted blue); 
(c) $N$ with: $x=1$, $\bar{N}=10$, $\lambda=10$ (solid green), $\lambda=1$ (dashed red), $\lambda=0.75$ (dotted blue); 
(d) $N$ with: $\bar{N}=10$, $\lambda=1$; $x=10$ (solid green), $x=5$ (dashed red), $x=1$ (dotted blue). We have  assumed unit values for all the other parameters.}
 \label{fig3}
\end{figure*}

Similarly, we can write the  pressure $P$ and    chemical potential $\mu$ for galaxies interacting through a nonlocal gravitational potential   as
\begin{align}
P&= -\biggl(\frac{\partial F}{\partial V}\biggr)_{N,T}
= \frac{NT}{V}\big(1-B_l\big),
\nonumber \\
\mu &= \biggl(\frac{\partial F}{\partial N}\biggr)_{V,T}
 =  T \ln (\rho T^{-3/2})  \\  
 &- T \ln \big(1+\alpha x\big)
 - T \frac{3}{2}\ln \big(\frac{2\pi m}{\Lambda^2}\big)- T B_l.\nonumber
\end{align}
The probability of finding $N$ galaxies can be written as
\begin{eqnarray}
F(N)=\frac{\sum_{i}e^{\frac{N\mu}{T}}e^{\frac{-U_{n}}{T}}}{Z_{G}(T,V,z)}=\frac{e^{\frac{N\mu}{T}}Z_{N}(V,T)}{Z_{G}(T,V,z)},
\end{eqnarray}
where $Z_{G}$ is the grand-partition function defined by
\begin{eqnarray}
Z_{G}(T,V,z)=\sum_{N=0}^{\infty}z^NZ_{N}(V,T).
\end{eqnarray}
and $z$ is the activity.
Thus, for a system of gravitationally interacting  galaxies in nonlocal gravity , we can write
\begin{align}
\exp \left[\frac{N\mu}{T}\right]&=\biggl(\frac{{\bar N}}{V}T^{-3/2}\biggr)^{N}\biggl(1+\frac{B_l}{(1-B_l)}\biggr)^{-N} \nonumber \\
&\cdot \exp\left[-NB_l\right]\biggl(\frac{2\pi m}{\Lambda^{2}}\biggr)^{-3N/2}.
\end{align}
Now the grand-partition function  can be written as
\begin{eqnarray}
\ln Z_{G}=\frac{PV}{T}=\bar N(1-B_l),
\end{eqnarray}
and the distribution function can be expressed as
\begin{align}\label{eq:pNV}
F(N) &= \frac{\bar{N}(1-B_l)}{N!}\left(NB_l+{\bar N}(1-B_l)\right)^{N-1} \nonumber \\
&\cdot \exp \left[-NB_l-\bar N(1-B_l)\right]\, .
\end{align}
In Fig. \ref{fig3},  we  analyze the behavior of the  distribution function for galaxies, and its dependence on different parameters used here. In general, the distribution function has the universal feature of having a maximum, while its functional shape changes, and depends on the considered parameters.

\subsection{Spatial correlation function}

It is well know that the interaction between different galaxies will cause correlation in their positions. The integral of the correlation function over a certain volume can be expressed in terms of the mean square fluctuation of the total number of galaxies in a given volume. Thus we can write
\begin{equation}
\int\xi dV=\frac{<(\Delta N)>^2}{\bar{N}}-1
\end{equation}
This can be further represented in terms of  thermodynamic quantities as
\begin{equation}
\int\xi dV=-\frac{NT}{V^2}\left(\frac{\partial V}{\partial P}\right)_T-1
\end{equation}
Taking the volume derivative and using the equation of state for pressure $P$,  we get
\begin{equation}
\xi=\frac{-NT}{V^2}\frac{\partial}{\partial V}\left(\frac{\partial V}{\partial P}\right)_T+2\frac{NT}{V^3}\left(\frac{\partial V}{\partial P}\right)_T
\end{equation}
The pressure can be written in terms of the  clustering parameter as
\begin{equation}
P=\frac{NT}{V}\big(1-B_l\big),
\end{equation}
We can write the change of volume with pressure at constant temperature
\begin{equation}
\left(\frac{\partial V}{\partial P}\right)_T=-\frac{V^2}{NT(1-B_l)^2}
\end{equation}
The expression for $\xi$ is
\begin{equation}
\xi=-2\frac{B_l}{V(1-B_l)^2}
\end{equation}
where ${\partial B_l}/{\partial V}=-({x}/{V})({dB_l}/{dx})=B_l(1-B_l)/{V}$ and $x=\beta\rho T^{-3}=\beta NT^{-3}/{V}$.
Now we can write 
\begin{equation}
\int\xi dV =\frac{B_l(2-B_l)}{(1-B_l)^2}
\end{equation}
The volume integral of correlation function is the quantity which can be compared with observational data in order to detect possible effects of nonlocal gravity.

\section{Observational Data Analysis}

We will test the  nonlocal  gravitational clustering with the cluster catalog in \citep{WenHan2012}, containing $132,684$ clusters of galaxies in the redshift range $0.05 \leq z < 0.8$ from the Sloan Digital Sky Survey III (SDSS-III). We will take advantage of this catalog because it has all the needed ingredients to analyze clustering properties of both galaxies and clusters (although an alternative more recent version is in \cite{WenHan2015}), using two different approaches.

\subsection{Galaxies}

The analysis of the clustering of galaxies will be conducted with a sort of ``smart'' version of the standard counts-in-cell procedure. In general, one would first need to define cells/volumes with a given size, and then to count the gravitational structures of interest within them. Here, our ``smart'' volumes will be the clusters identified in the SDSS-III survey, and we will count the galaxies which  are within each of them. From this procedure, we will be able to calculate the observed distribution function $F(N)$ and to compare it with the theoretical expectation in  nonlocal  gravity as given by Eq.~(\ref{eq:pNV}). The ``smartness'' in this procedure is that the cells/volumes we are going to analyze are physically realized systems, structures which exist and have been identified, contrarily to the standard way of proceeding, where the existence of clusters and interacting systems of galaxies is not assured neither verified. On the other hand, performing the count in this way, we are going to inevitably miss voids, which instead indirectly retain some information about the clustering properties.

All the needed data are provided by the catalogue in \citep{WenHan2012}: the volume of the cells or, equivalently, the radius of the clusters, $r_{200}$ in Mpc, is defined as the radius within which the mean density of a cluster is $200$ times the critical density of the Universe at the same redshift  and
 $N_{200}$ is the number of member galaxy candidates within $r_{200}$ of each cluster. The only caveat to take in mind is that the radii of the cells we are considering, i.e. the $r_{200}$, are much smaller than the scales where the quasi-equilibrium clustering should be more effective \citep{Yang11,Yang12}; we will discuss  this point when presenting our results.

We have divided the full catalogue in groups by redshift, with bin widths of $\Delta z = 0.05$, and by radius, with bin widths of $\Delta r_{200} = 0.1$ Mpc, and we have eventually selected only the groups with a sufficiently large number of clusters to enable a strong and reliable statistical analysis. The final groups with which we will work will have radii in the three ranges, $0.8 < r_{200} < 0.9$, $0.9 < r_{200} < 1.0$ and $1.0 < r_{200} < 1.1$ Mpc, and redshifts in the interval $0.05 \leq z \leq 0.65$. We need to remind here that the catalogue is complete only up to a redshift $z \sim 0.42$, for clusters with an estimated mass $M_{200}> 1 \cdot 10^{14}$ $M_{\odot}$; at higher redshift, a bias toward larger clusters (larger masses) and possibly to higher counts-in-cell is possible \citep{WenHan2012}.

In Table~\ref{tab:results_galaxies},  we show the results derived from using Eq.~(\ref{eq:pNV}) with the chosen data. The parameters involved in the analysis of Eq.~(\ref{eq:pNV}) have been rewritten as:
\begin{itemize}
  \item[-] $R=r_{200}$, the radius of the cell (cluster), which enters Eq.~(\ref{eq:pNV}) through Eq.~(\ref{eq:alpha}). We have assumed it to be constant within the three bins we have identified; in the following, it will act as a scaling factor for some parameters;
  \item[-] $N=N_{200}$, the number of galaxies within each cell (cluster);
  \item[-] the clustering parameter $b$, defined as
  \begin{equation}
  b = \frac{x}{1+x}\, ,
  \end{equation}
  and which enters Eq.~(\ref{eq:pNV}) through:
        \begin{equation}
        B_{l} = \frac{b\, \alpha}{1+b(\alpha-1)} \; ,
        \end{equation}
        with $\alpha$ given by Eq.~(\ref{eq:alpha});
  \item[-] the dimensionless softening parameter, $\tilde{\epsilon} = \epsilon/R$;
  \item[-] the dimensionless  nonlocal  characteristic length, $\tilde\lambda = \lambda/R$.
\end{itemize}
The only  parameter which fully characterizes the nonlocal gravity model we are focusing on is thus $\tilde{\lambda}$. All other parameters are standard in the sense that they would appear also whether standard GR would be considered. In Table~\ref{tab:results_galaxies} we also report  the number of clusters in each redshift bin, $N_{cl}$, although this is not a fitting parameter.

When it comes to fit the data, we have defined and compared results from two different statistical tools. We have performed a least-square minimization using the $\chi^2$ defined as
\begin{equation}\label{eq:chils}
\chi^{2}_{ls} = \sum_{i}\left[F^{theo}(N_{i}) - F^{obs}(N_{i})\right]^2,
\end{equation}
where $F^{obs}(N_{i})$ is the real counts-in-cell distribution extracted from the data, and $F^{theo}(N_{i})$ is the theoretical counts-in-cell calculated from the nonlocal gravity model by using Eq.~(\ref{eq:pNV}). The index $i$ derives from the fact that in each radius bin, the clusters  have a variable amount of galaxies within them, ranging from some minimal number $N_{min}$ to a maximal one $N_{max}$. Thus, the index $i$ selects the finite natural values of this range, where we evaluate the distribution function. We have also defined a second $\chi^2$ as
\begin{equation}\label{eq:chijk}
\chi^{2}_{jk} = \sum_{i} \frac{\left[F^{theo}(N_{i}) - F^{obs}(N_{i})\right]^2}{\sigma^{2}_{i}},
\end{equation}
where the $\sigma_i$ are the errors on $F^{obs}(N_{i})$ which we have derived from a jackknife-like procedure as exposed below. 

For each redshift and radius bin: 
\begin{enumerate}
 \item we cut a variable fraction $\mathcal{F}$ (randomly selected in the range $[10,90]\%$) from the total bin population; 
 \item for each cut sample we derive the counts-in-cell distribution function $\sim 50$ times;
 \item we thus obtain a distribution of $F^{obs}(N_{i})$ from the $\sim 50$ sets of $N_i$;
 \item from each distribution, which looks very close to a standard Gaussian, we derive the standard deviation and assume such value as the error $\sigma_i$. 
\end{enumerate}
The data points and the errors on the distribution function $F^{obs}(N_{i})$ which we finally obtain are shown respectively as black dots and bars in Figs.~(\ref{fig:R5})~-~(\ref{fig:R6})~-~(\ref{fig:R7}). The best fitting $F(N)$ distributions obtained from the minimization of $\chi^2_{ls}$ are shown in red; those ones derived from $\chi^2_{jk}$ are in green.

The minimization of the defined $\chi^2$ is performed by using a Monte Carlo Markov Chain (MCMC) approach, running chains with $10^6$ points and using the uninformative and very general priors: $\bar{N}\geq0$; $0\leq b \leq 1$; $0\leq \tilde{\epsilon} \leq 1$; $\tilde{\lambda}>0$.

\subsection{Clusters}

\begin{figure*}
\centering
\includegraphics[width=0.5\textwidth]{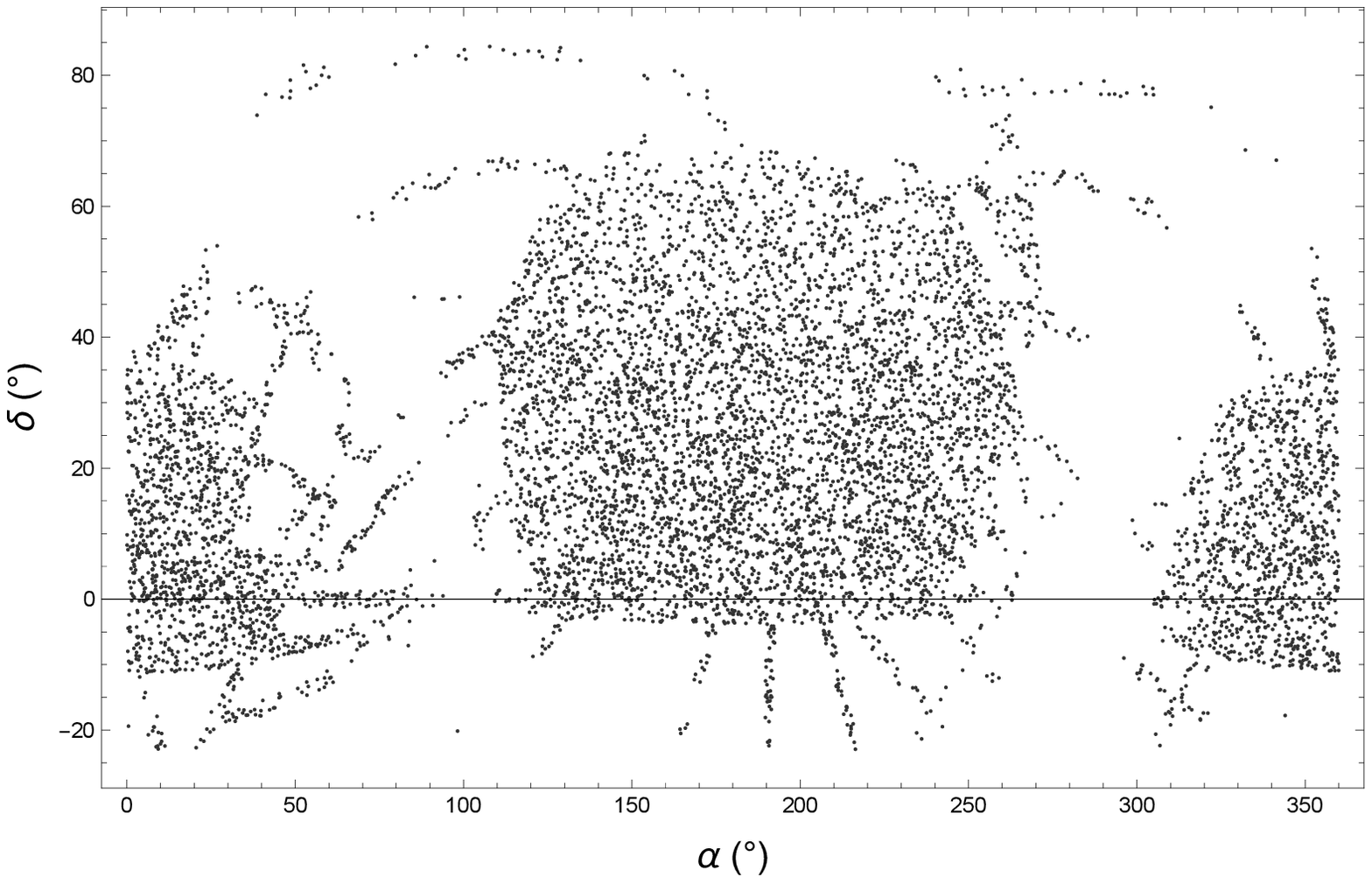}\\
~~~\\
\includegraphics[width=0.3\textwidth]{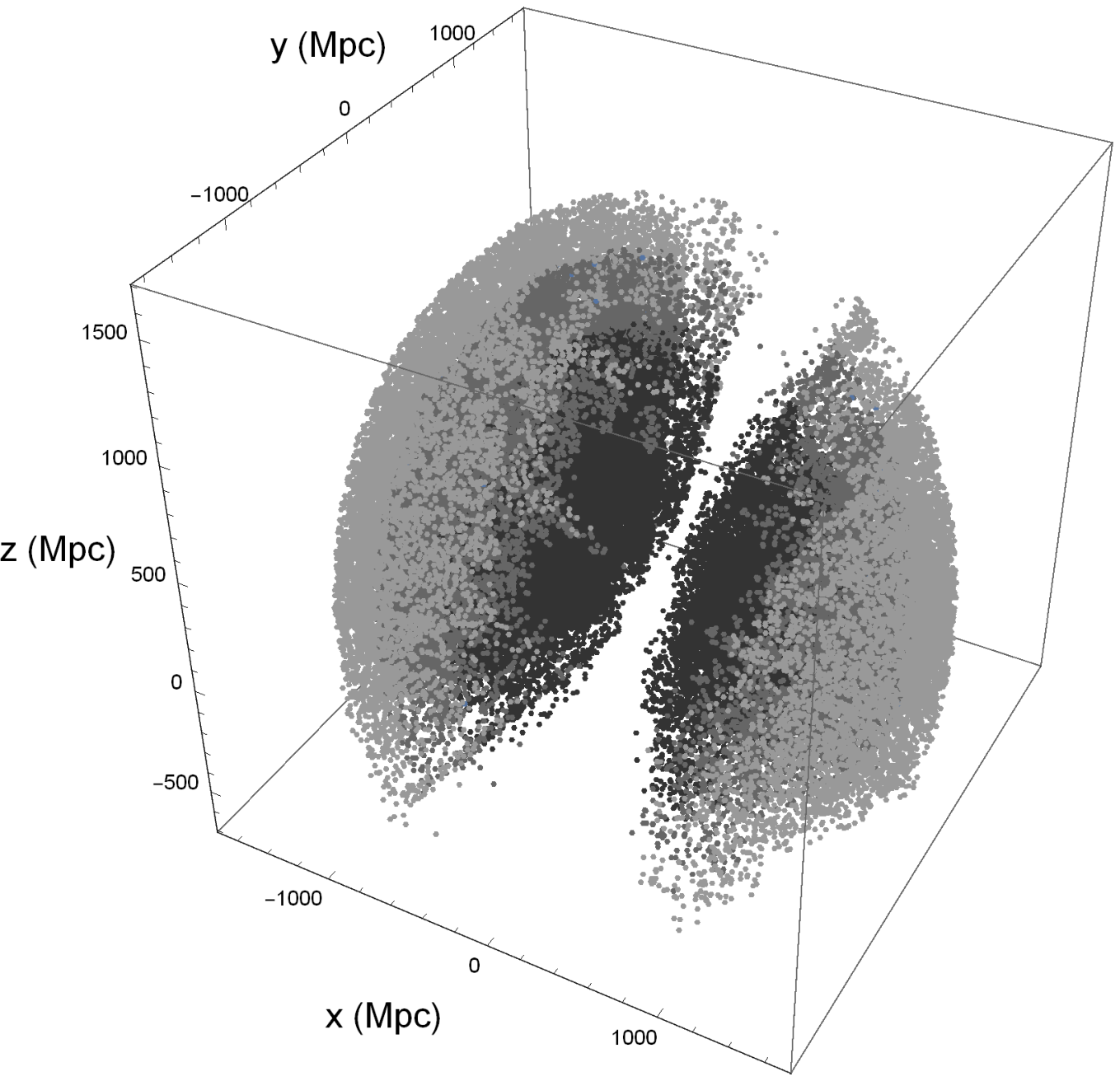}~~~~~~\includegraphics[width=0.3\textwidth]{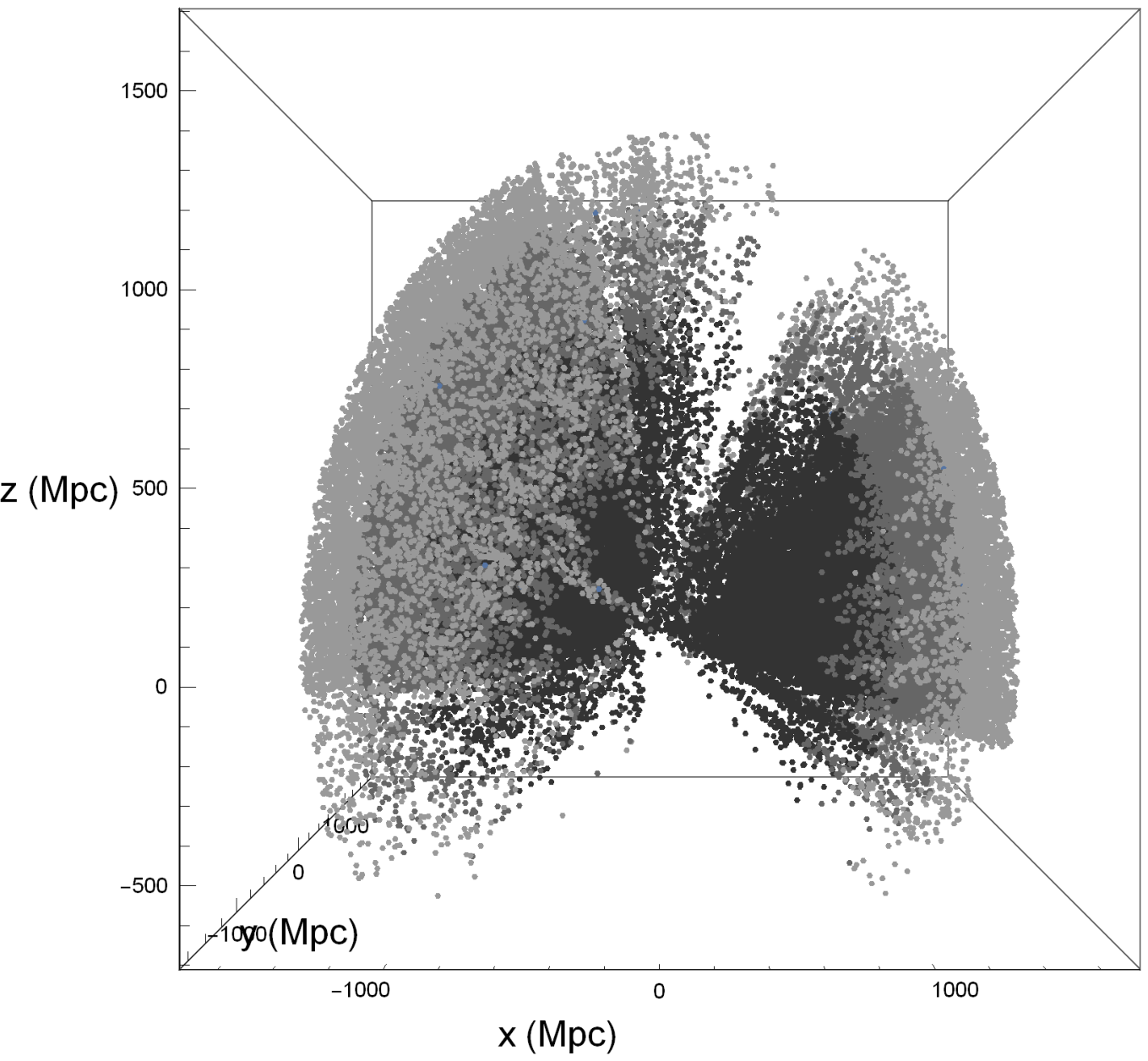}~~~~~~\includegraphics[width=0.3\textwidth]{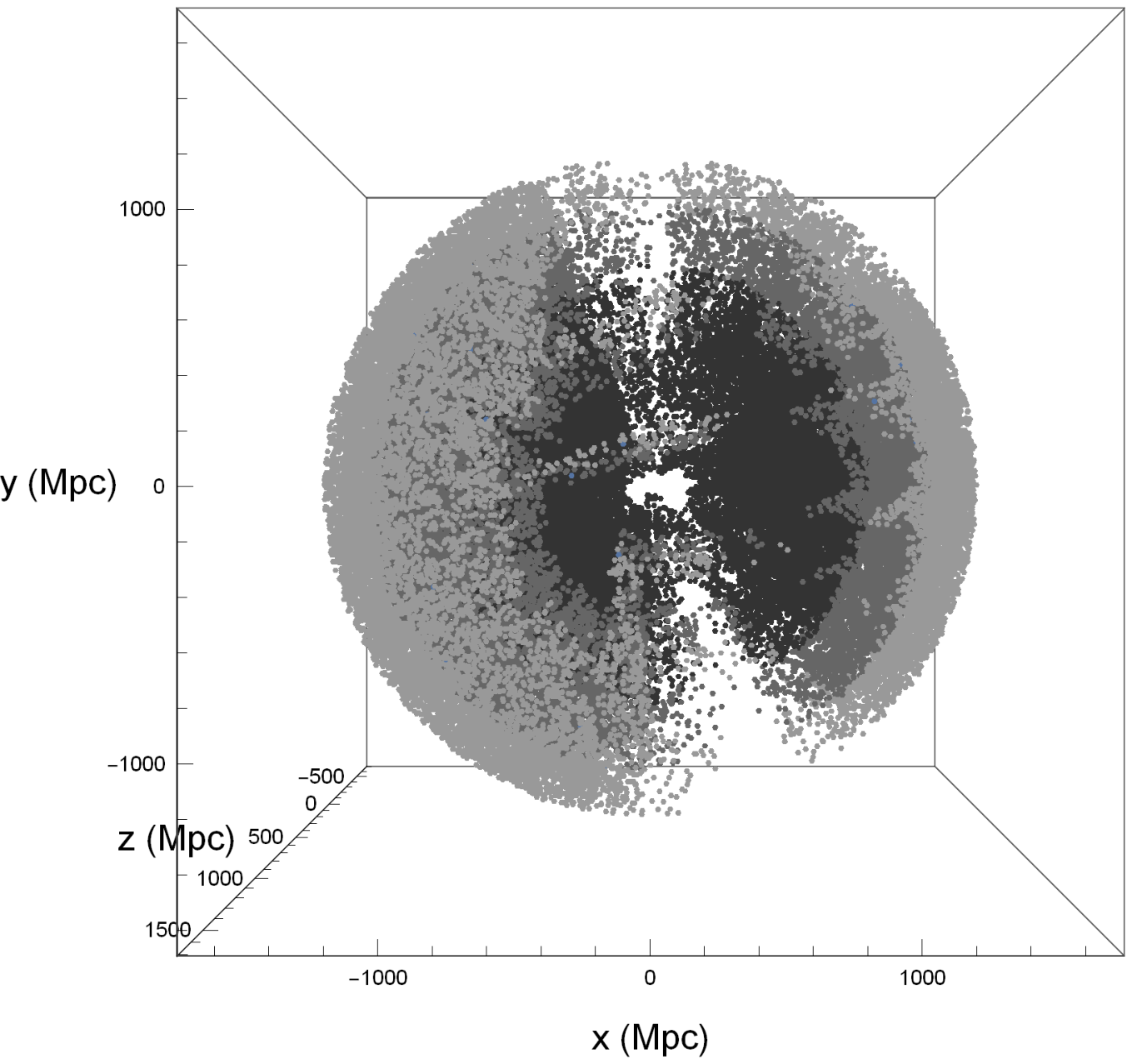}\\
\caption{SDSS survey area. \textit{(Top panel.)} Full clusters catalog in Celestial coordinates, right ascension $\alpha$ and declination $\delta$. \textit{(Bottom panel.)} Full clusters catalog in physical distances from three different viewpoints. The different colors represent the three redshift bins we have described in the text, with: $0.05 \leq z < 0.281$ as dark grey; $0.281 \leq z < 0.361$ as middle grey; and $0.361 \leq z < 0.42$ as light grey.}\label{fig:SDSS}
\end{figure*}

Concerning the count-in-cells where the counted objects within volumes are  the clusters of galaxies, we follow a procedure which is very similar to what is  described in \citep{Yang11}. In order to be as much as clear as possible, we are going to enumerate all the steps in the following list. 

The initial set up consists in the coming steps:
\begin{enumerate}
  \item[(1a)] we choose the size of the spherical cells/volumes we are going to analyze. We have selected four different physical lengths, $R=10$, $20$, $30$, and $40$ Mpc;
  \item [(2a)] we divide the clusters in three groups by redshift, $0.05\leq z < 0.281$, $0.281\leq z < 0.361$ and $0.361\leq z < 0.42$. These values have been chosen in order to span approximately the same comoving volume in each group: assuming a baseline \textit{Planck} cosmology \citep{Aghanim:2018eyx}, with $H_0 = 67.51$ km s$^{-1}$ Mpc$^{-1}$ and $\Omega_m=0.3121$, the volume is $\approx 6.5 \cdot 10^9$ Mpc$^{3}$. The number of clusters from the SDSS catalog falling into each group are, respectively, $37686$, $24208$ and $20634$ (the total sums up to $82528$, because we are taking only clusters with $z \leq 0.42$, for which the catalogue is complete);
  \item[(3a)] all clusters in the catalog are provided with redshift $z$ and celestial coordinates, the $J2000.0$ right ascension $\beta$ and declination $\delta$. These quantities are converted into physical Cartesian coordinates using the standard formulae,
      \begin{eqnarray}
      x &=& D_{C}(z) \cos \delta \cos \beta\, , \nonumber \\
      y &=& D_{C}(z) \cos \delta \sin \beta\, , \\
      z &=& D_{C}(z) \sin \delta \cos \beta\, , \nonumber \\
      \end{eqnarray}
        where $D_{C}$ is the comoving distance calculated assuming the same fiducial cosmology as in the previous step. In Fig.~\ref{fig:SDSS} we  show both the angular projection of the SDSS survey area (top panel), and the ``physical length'' distribution of the catalogue, with the three bins shown in different colors (bottom panel);
  \item[(4a)] we define a Cartesian grid covering the SDSS survey area. Each point in this grid will be the center of a cell which will be used for the count-in-cell procedure. In order to guarantee a full coverage of the area, considering that  the volumes are spheres, the distance between each point (i.e. each center of each cell) is $\sqrt{2} R$;
  \item[(4a)] the SDSS survey does not cover the sky in a uniform way; thus, we are not going to include in our analysis all the points from the previously defined Cartesian grid. Actually, converting the Cartesian coordinates of each point back to celestial coordinates, we only select those ones which fall in the range defined by the initial cluster catalog. This step will acquire more significance as further steps in our procedure will be highlighted in the next points.
\end{enumerate}
Having defined this initial set up, we have performed the following operations on/with it:
\begin{enumerate}
  \item[(1b)] we randomize the grid defined in the previous point $(3a)$, by shifting its origin by a random number in the interval $[0,\sqrt{2}R]$. Actually, we apply this shift independently to each Cartesian axis of the grid;
  \item[(2b)] after each shift, we only retain cells which satisfy the condition described in the previous point $(5a)$. This assures us that any cell which should fall out of the SDSS survey will not be considered in the following analysis;
  \item[(3b)] additionally, we apply a jackknife cut of $10\%$ of cells selected randomly;
  \item[(4b)] the previous steps, $(1b)-(2b)-(3b)$, are realized $100$ times, each time having a different shift in the origin of the grid system (point $(1b)$), a possibly different number of cells falling out of the SDSS survey are (point $(2b)$), and a different set of cells cut from point $(3b)$.
\end{enumerate}
After all these steps are performed, we eventually have at our disposal the grid of centers of cells/volumes, for each cell size $R$ provided at point $(1a)$, and divided in the three subgroups following the redshift criterion  described at point $(2a)$, which will serve to perform the count-in-cell with the cluster catalog data. Basically, for each point in the grid, we will retain only the clusters whose distance from a given grid point (i.e. cell center) will be lower than the size $R$ taken into account. Because of point $(4b)$ above, we basically have $100$ different grid sets to compare with the original catalogue, namely, we will have $100$ $F(N)$ spanning the ranges $[N_{min},N_{max}]$ where, as before, $N_{min}$ and $N_{max}$ are the minimum and maximum numbers of objects (clusters) within each cell/volume.

The median of such ensemble of $100$ values for $F(N)$ will serve as observational estimation for this quantity in the $\chi_{ls}$ analysis, as defined in Eq.~(\ref{eq:chils}). Instead, the median and the standard deviation, assumed as error $\sigma_{i}$, will be used when performing the $\chi_{jk}$ analysis, as defined in Eq.~(\ref{eq:chijk}).

\subsection{Results}

The complete outcomes of our statistical analysis are presented in Table~\ref{tab:results_galaxies} for galaxies and Table~\ref{tab:results_clusters} for clusters of galaxies. As a first comment, valid for both cases, let us notice that the softening parameter $\tilde{\epsilon}$ is not present in the tables because it is basically unconstrained, exhibiting a uniform distribution all over the given prior range $[0,1]$. Thus, we can infer that its role in the fit is marginal.

\subsubsection{Galaxies}

For what concerns galaxies, from Table~\ref{tab:results_galaxies} we can see how the parameter $\bar{N}$ is very well constrained, and there is no difference between using $\chi^2_{ls}$ or $\chi^2_{jk}$. This statement is valid whatever radius $R$ and redshift bin is considered. Actually, we must remember that the catalog is complete only up to $z=0.42$, after which we do have a bias toward smaller clusters, namely, clusters with a lower $\bar{N}$. And this is effectively seen in the tables, where $\bar{N}$ rises slightly with redshift, becomes practically constant, and then starts to decrease for $z>0.45$.

The same consideration is basically true also for the clustering parameter $b$, even if in this case  we can see how there is a trend with varying $R$, while no peculiar change with redshift $z$ can be detected. In general, this parameter lies in the range $[0.3,0.4]$ for all cases; however, while for the smaller $R$ we can see how it is very well constrained, for growing $R$, we are able to put only upper limits on it, which means that also small values of $b$ are compatible with data. As described in \citep{Yang12}, we know that smaller values of $b$ correspond to clusters which have not yet fully virialized. It is worth noticing   that such trend is on the opposite side with respect to \citep{10,20,Yang12} where smaller cells volumes correspond to smaller values of $b$; although we must also stress that in \citep{10,20,Yang12}, the considered physical volumes are much larger than those we have defined here as well as the redshift bins, and the number of objects in each bin is much lower. In our case the bins are much finer, both in radius and in redshift.

For what concerns the only fully characterizing parameter of the nonlocal gravitational potential, i.e.  $\tilde{\lambda}$, we can see how it is remarkably very well constrained in a  specific and well defined range. Statistically speaking, we can see how it approximately  lies in the range $[0.4,0.5]$, which, in physical units, corresponds roughly to $[0.32,0.55]$ Mpc. We can detect a slight decrease for varying redshift, with smaller values at larger $z$; while it seems to be more pronounced a trend with varying $R$, with larger $\tilde{\lambda}$ for larger clusters. It is quite striking to have such values for this parameter, which are actually totally consistent with the size of  gravitational structures/volumes under scrutiny. At the same time, they are also very different from a stable GR limit, which would be achieved when $\lambda \rightarrow \infty$. Actually, from Eq.~\ref{eq:nonlocal_pot}, one can see how it would be possible to recover an \textit{unstable} GR-like limit even for $\lambda \rightarrow r_{ij}$. 

\begin{figure*}
\includegraphics[width=\textwidth]{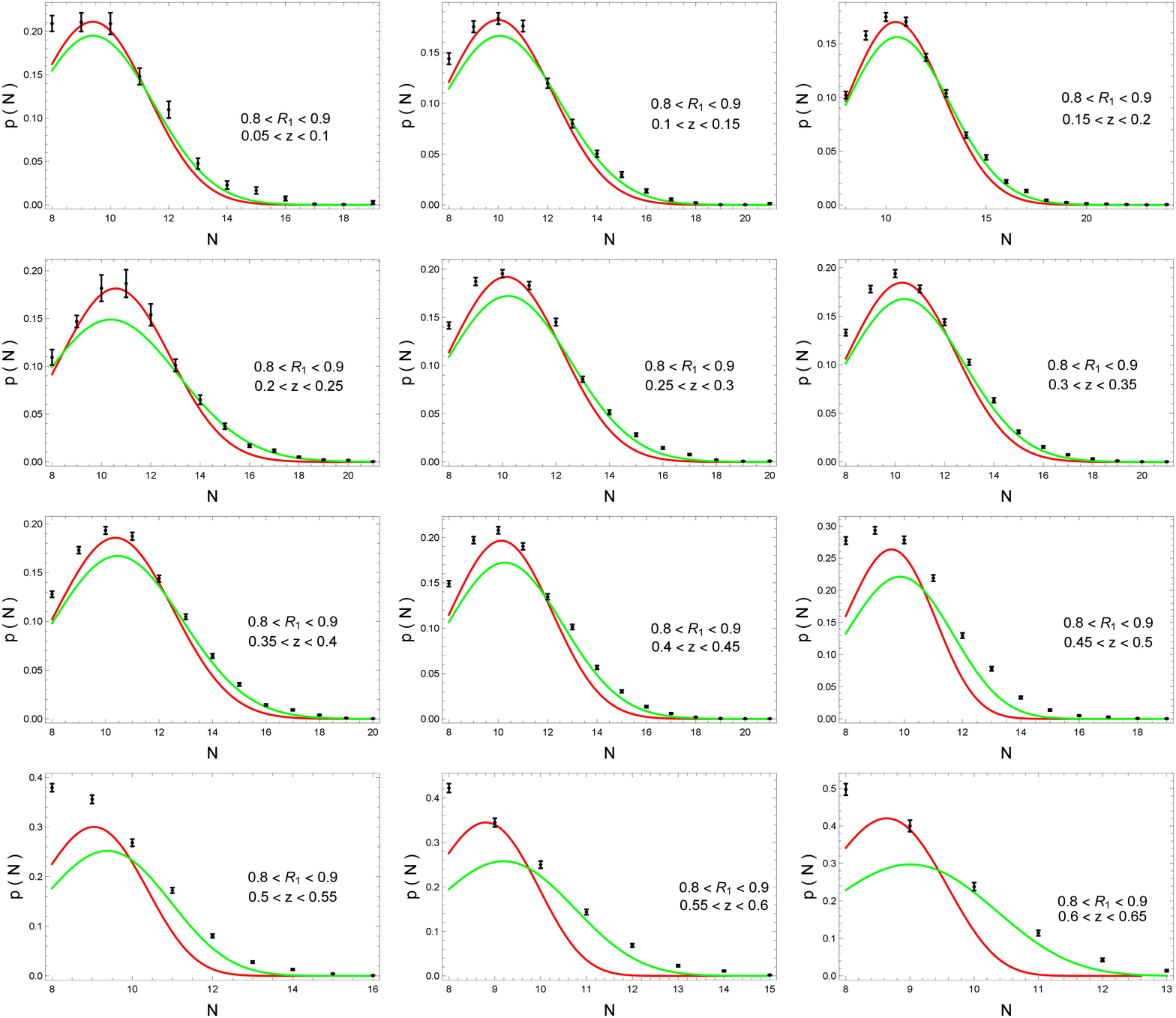}
\caption{Comparison between data and theoretical expectation Eq.~(\ref{eq:pNV}) for cell size bin $0.8 < R < 0.9$ $Mpc$. Black points: data; black bars: jackknife-like observational errors. Solid red line: best fit from minimization of $\chi_{ls}^{2}$; solid green line: best fit from minimization of $\chi_{jk}^{2}$.}\label{fig:R5}
\end{figure*}

\begin{figure*}
\includegraphics[width=\textwidth]{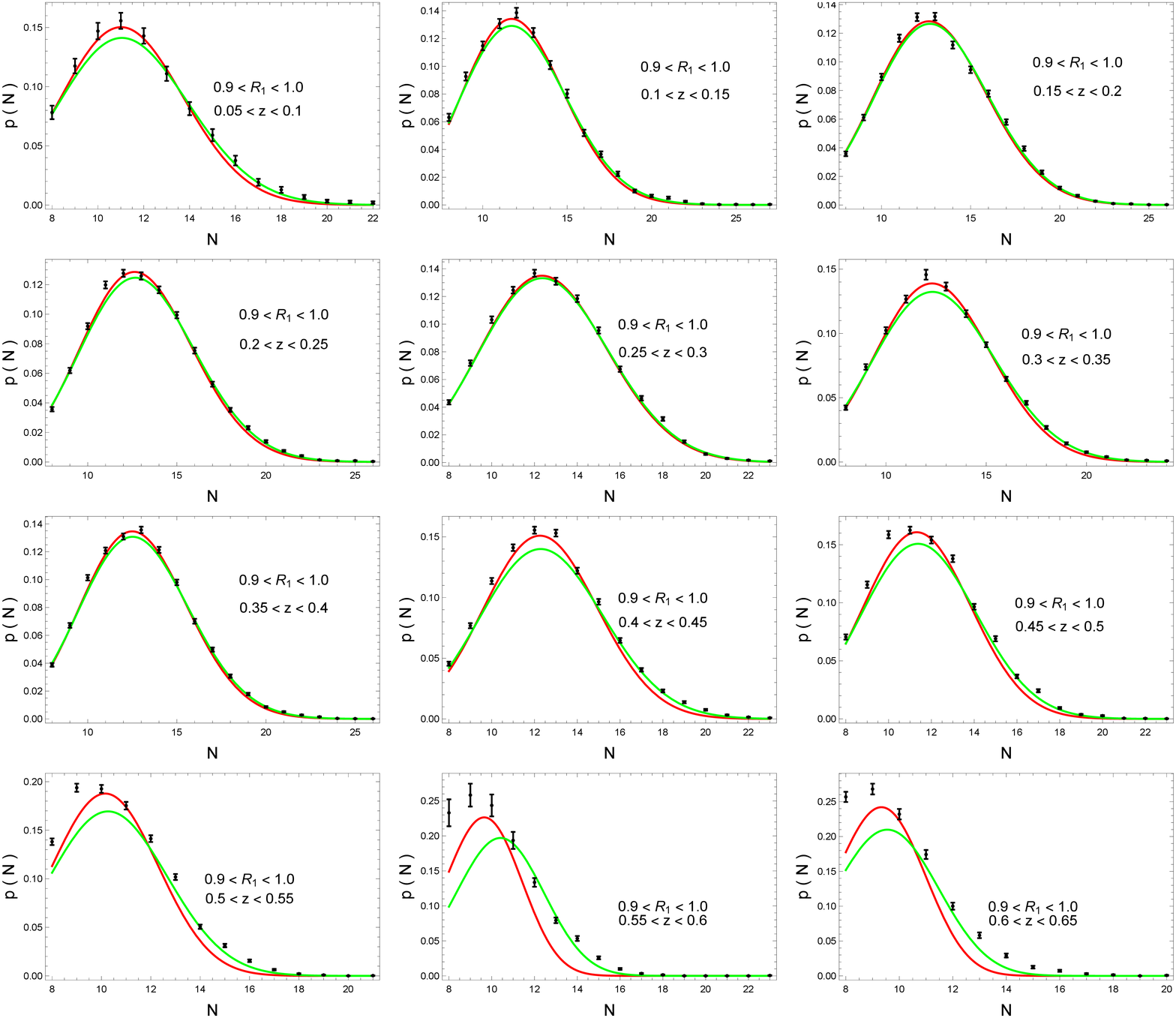}
\caption{Comparison between data and theoretical expectation Eq.~(\ref{eq:pNV}) for cell size bin $0.9 < R < 1.0$ $Mpc$. Black points: data; black bars: jackknife-like observational errors. Solid red line: best fit from minimization of $\chi_{ls}^{2}$; solid green line: best fit from minimization of $\chi_{jk}^{2}$.}\label{fig:R6}
\end{figure*}

\begin{figure*}
\includegraphics[width=\textwidth]{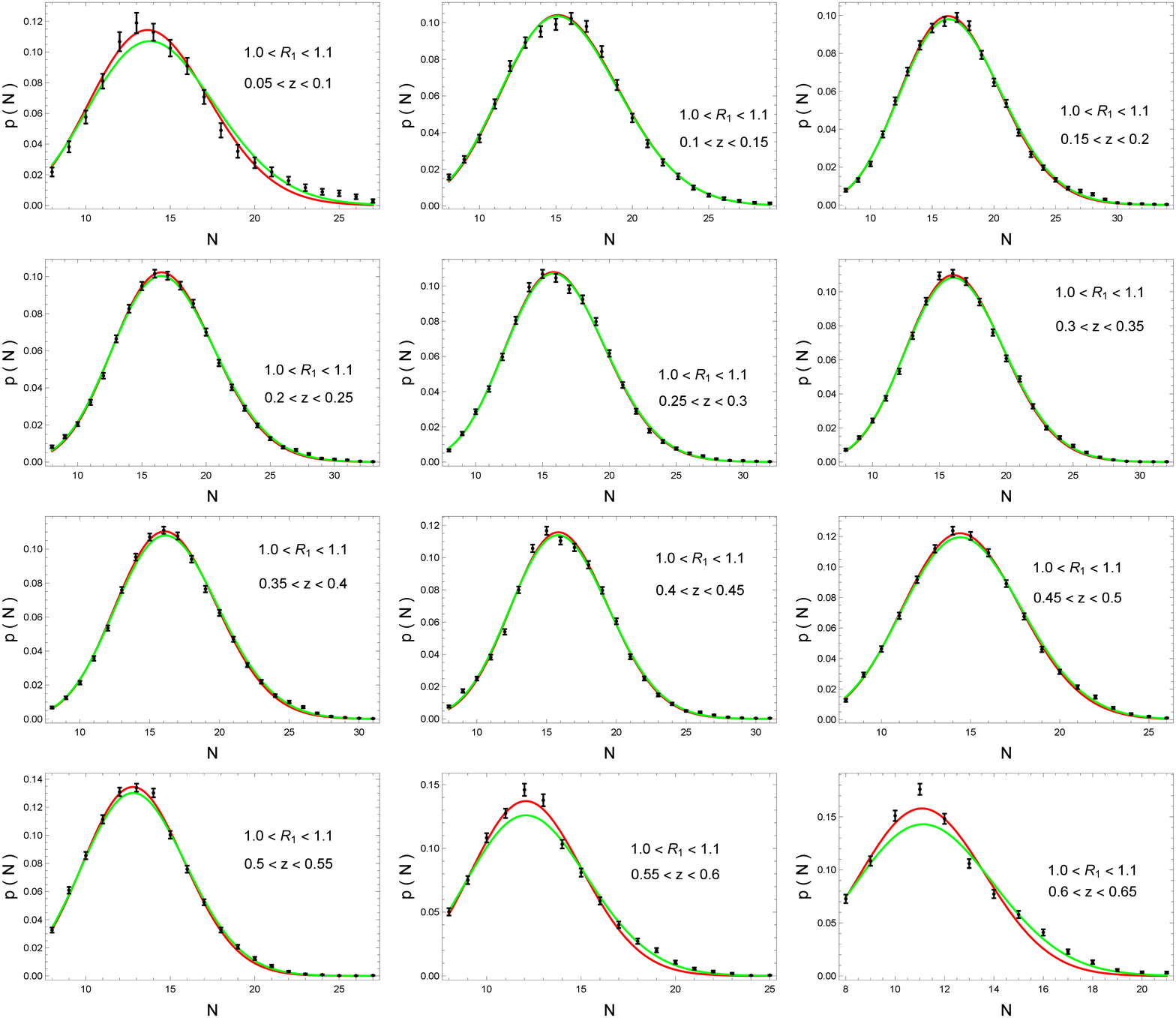}
\caption{Comparison between data and theoretical expectation Eq.~(\ref{eq:pNV}) for cell size bin $1.0 < R < 1.1$ $Mpc$. Black points: data; black bars: jackknife-like observational errors. Solid red line: best fit from minimization of $\chi_{ls}^{2}$; solid green line: best fit from minimization of $\chi_{jk}^{2}$.}\label{fig:R7}
\end{figure*}

Finally, from Figs.~\ref{fig:R5}~-~\ref{fig:R6}~-~\ref{fig:R7},  we can also visually check the quality of the fits: the larger is the cell volume, the better is the fit.

We must stress here an important point: these results just show that a logarithmic correction to the Newtonian gravity might be consistent with the clustering of galaxies. But we cannot state at all that this correction is consistent with the full clustering process at these same scales, because, from the catalogue we have used, we  have no information on voids, which should be taken into account. This is a further check which should be performed in the future.

\subsubsection{Clusters}

What we can check is how the full clustering process on larger scales behaves. As we have stated in the previous sections, if we move to consider the clustering of clusters of galaxies, the catalogue in hand does allow us to properly take into account voids. Although, in this case, results are a bit more fuzzy.

First of all, we focus on $\bar{N}$. We can see here a trend both in redshift and in volume size $R$: $\bar{N}$ decreases with redshift and increases with $R$. The latter correspondence is of course expected, as large cells can contain more clusters; the former one is also somehow expected, because we might predict an intrinsic lack of clusters at higher redshifts, as they are caught in an epoch of formation, while a larger number of them is observable at smaller redshifts. Notice that our bins all have the same comoving volumes, so that any geometry-related issue should not be effective in this case.

The behavior of the clustering parameter $b$ is now fully consistent with literature \citep{10,20,Yang12}, with smaller values for smaller volumes, as well as for higher redshifts.

The nonlocal characteristic length $\tilde{\lambda}$, instead, in this case is quite unconstrained. We are unable to put strong boundaries on it and in Table~\ref{tab:results_clusters} we can only assess upper limits on it. Although we must warn that such results are statistically weak, with the posteriors being quite irregular in most of the cases. This result can only be interpreted as a substantial negligible role of the parameter $\lambda$ in the fitting procedure. Thus, we should conclude that on such large scales there is no evidence for a detectable logarithmic correction to the standard Newtonian potential.

{\renewcommand{\tabcolsep}{1.5mm}
{\renewcommand{\arraystretch}{2.}
\begin{table*}
\begin{minipage}{\textwidth}
\caption{Galaxies: results from fitting Eq.~(\ref{eq:pNV}) with data from the cluster catalog from \citep{WenHan2012}. Data and fits are divided in redshift bins (first column) and cell radius bin (as indicated by $R$ intervals). $N_{cl}$ is not a fitting parameter, but the number of clusters in each redshift and radius bin. All the fitting parameters are described in the text.}\label{tab:results_galaxies}
\centering
\resizebox*{\textwidth}{!}{
\begin{tabular}{ccccccccccccc}
\hline
 & \multicolumn{4}{c}{$0.8<R<0.9$ Mpc} & \multicolumn{4}{c}{$0.9<R<1.0$ Mpc} & \multicolumn{4}{c}{$1.0<R<1.1$ Mpc}  \\
            \cmidrule(lr){2-5} \cmidrule(lr){6-9} \cmidrule(lr){10-13}
$z$          & $N_{cl}$ & $\bar{N}$ & $b$ & $\tilde{\lambda}$
             & $N_{cl}$ & $\bar{N}$ & $b$ & $\tilde{\lambda}$
             & $N_{cl}$ & $\bar{N}$ & $b$ & $\tilde{\lambda}$ \\
\hline
\multicolumn{13}{c}{$\chi^{2}_{ls}$} \\
$[0.05,0.1]$ &
$291$ &  $9.34^{+0.09}_{-0.09}$ & $0.35^{+0.22}_{-0.17}$ & $0.39^{+0.12}_{-0.10}$ &
$390$ & $11.16^{+0.13}_{-0.13}$ & $0.29^{+0.24}_{-0.16}$ & $0.42^{+0.12}_{-0.10}$ &
$298$ & $14.04^{+0.11}_{-0.11}$ & $<0.30$ & $0.46^{+0.26}_{-0.34}$ \\
$[0.1,0.15]$ &
$1052$ & $10.00^{+0.11}_{-0.11}$ & $0.32^{+0.22}_{-0.16}$ & $0.40^{+0.12}_{-0.10}$ &
$1526$ & $11.96^{+0.16}_{-0.16}$ & $<0.36$ & $0.44^{+0.11}_{-0.10}$ &
$949$  & $15.59^{+0.04}_{-0.04}$ & $<0.26$ & $0.48^{+0.12}_{-0.10}$ \\
$[0.15,0.2]$ &
$2343$ & $10.54^{+0.11}_{-0.11}$ & $0.32^{+0.22}_{-0.17}$ & $0.40^{+0.12}_{-0.10}$ &
$2919$ & $12.93^{+0.17}_{-0.16}$ & $<0.37$ & $0.44^{+0.12}_{-0.10}$ &
$1826$ & $16.75^{+0.03}_{-0.03}$ & $<0.23$ & $0.49^{+0.12}_{-0.10}$ \\
$[0.2,0.25]$ &
$3069$ & $10.58^{+0.10}_{-0.10}$ & $0.34^{+0.24}_{-0.18}$ & $0.40^{+0.12}_{-0.10}$ &
$3688$ & $12.90^{+0.16}_{-0.16}$ & $<0.35$ & $0.44^{+0.12}_{-0.10}$ &
$2398$ & $16.92^{+0.07}_{-0.07}$ & $<0.30$ & $0.47^{+0.12}_{-0.10}$ \\
$[0.25,0.3]$ &
$3318$ & $10.12^{+0.09}_{-0.10}$ & $0.34^{+0.22}_{-0.17}$ & $0.40^{+0.12}_{-0.10}$ &
$3982$ & $12.56^{+0.15}_{-0.15}$ & $<0.39$ & $0.42^{+0.12}_{-0.10}$ &
$2443$ & $16.17^{+0.11}_{-0.12}$ & $<0.33$ & $0.45^{+0.12}_{-0.10}$ \\
$[0.3,0.35]$ &
$4024$ & $10.28^{+0.05}_{-0.05}$ & $0.33^{+0.23}_{-0.16}$ & $0.39^{+0.12}_{-0.09}$ &
$5055$ & $12.47^{+0.14}_{-0.14}$ & $<0.39$ & $0.42^{+0.12}_{-0.10}$ &
$2958$ & $16.35^{+0.11}_{-0.11}$ & $<0.34$ & $0.45^{+0.12}_{-0.10}$ \\
$[0.35,0.4]$ &
$4379$ & $10.35^{+0.10}_{-0.10}$ & $0.34^{+0.23}_{-0.17}$ & $0.40^{+0.12}_{-0.10}$ &
$5402$ & $12.70^{+0.15}_{-0.15}$ & $<0.39$ & $0.43^{+0.12}_{-0.10}$ &
$3208$ & $16.37^{+0.11}_{-0.12}$ & $<0.35$ & $0.44^{+0.12}_{-0.10}$ \\
$[0.4,0.45]$ &
$4925$ & $10.09^{+0.09}_{-0.09}$ & $0.35^{+0.22}_{-0.17}$ & $0.39^{+0.12}_{-0.10}$ &
$5819$ & $12.35^{+0.13}_{-0.13}$ & $<0.41$ & $0.41^{+0.12}_{-0.10}$ &
$3289$ & $16.09^{+0.09}_{-0.09}$ & $<0.37$ & $0.43^{+0.11}_{-0.10}$ \\
$[0.45,0.5]$ &
$3982$ & $9.43^{+0.06}_{-0.06}$ & $0.36^{+0.23}_{-0.17}$ & $0.37^{+0.11}_{-0.10}$ &
$4813$ & $11.38^{+0.12}_{-0.12}$ & $<0.44$ & $0.41^{+0.13}_{-0.10}$ &
$2660$ & $14.64^{+0.18}_{-0.18}$ & $<0.37$ & $0.44^{+0.11}_{-0.10}$ \\
$[0.5,0.55]$ &
$2838$ & $8.88^{+0.05}_{-0.05}$ & $0.36^{+0.23}_{-0.17}$ & $0.35^{+0.14}_{-0.10}$ &
$3631$ & $10.15^{+0.10}_{-0.10}$ & $0.33^{+0.22}_{-0.17}$ & $0.39^{+0.11}_{-0.10}$ &
$2009$ & $12.98^{+0.15}_{-0.15}$ & $<0.38$ & $0.42^{+0.11}_{-0.10}$ \\
$[0.55,0.6]$ &
$1971$ & $8.64^{+0.04}_{-0.04}$ & $0.40^{+0.22}_{-0.19}$ & $0.37^{+0.12}_{-0.11}$ &
$2614$ & $9.57^{+0.07}_{-0.07}$ & $0.38^{+0.22}_{-0.18}$ & $0.39^{+0.12}_{-0.10}$ &
$1538$ & $12.27^{+0.15}_{-0.15}$ & $<0.40$ & $0.43^{+0.12}_{-0.10}$ \\
$[0.6,0.65]$ &
$1046$ & $8.47^{+0.03}_{-0.03}$ & $0.41^{+0.21}_{-0.29}$ & $0.36^{+0.12}_{-0.10}$ &
$1586$ & $9.23^{+0.07}_{-0.07}$ & $0.38^{+0.22}_{-0.18}$ & $0.38^{+0.12}_{-0.10}$ &
$977$ & $11.17^{+0.13}_{-0.13}$ & $<0.43$ & $0.42^{+0.11}_{-0.10}$ \\
\multicolumn{13}{c}{$\chi^{2}_{jk}$} \\
$[0.05,0.1]$ &
$291$ & $9.39^{+0.08}_{-0.08}$ & $0.34^{+0.22}_{-0.17}$ & $0.40^{+0.12}_{-0.10}$ &
$390$ & $11.26^{+0.07}_{-0.07}$ & $<0.38$ & $0.42^{+0.12}_{-0.10}$ &
$298$ & $14.28^{+0.08}_{-0.08}$ & $<0.02$ & $<60471$ \\
$[0.1,0.15]$ &
$1052$ & $10.15^{+0.05}_{-0.05}$ & $0.32^{+0.22}_{-0.16}$ & $0.41^{+0.11}_{-0.09}$ &
$1526$ & $12.00^{+0.04}_{-0.04}$ & $<0.34$ & $0.44^{+0.11}_{-0.10}$ &
$949$ & $15.55^{+0.04}_{-0.04}$ & $<0.23$ & $0.50^{+0.12}_{-0.10}$ \\
$[0.15,0.2]$ &
$2343$ & $10.69^{+0.03}_{-0.04}$ & $0.30^{+0.23}_{-0.16}$ & $0.42^{+0.12}_{-0.10}$ &
$2919$ & $12.99^{+0.03}_{-0.03}$ & $<0.40$ & $0.44^{+0.11}_{-0.08}$ &
$1826$ & $16.82^{+0.04}_{-0.04}$ & $<0.15$ & $0.54^{+0.13}_{-0.12}$ \\
$[0.2,0.25]$ &
$3069$ & $10.56^{+0.07}_{-0.08}$ & $<0.37$ & $0.42^{+0.12}_{-0.10}$ &
$3688$ & $12.97^{+0.03}_{-0.03}$ & $<0.35$ & $0.44^{+0.12}_{-0.10}$ &
$2398$ & $16.94^{+0.03}_{-0.03}$ & $<0.26$ & $0.47^{+0.12}_{-0.10}$ \\
$[0.25,0.3]$ &
$3318$ & $10.26^{+0.03}_{-0.03}$ & $0.34^{+0.24}_{-0.17}$ & $0.41^{+0.12}_{-0.09}$ &
$3982$ & $12.59^{+0.03}_{-0.03}$ & $<0.37$ & $0.42^{+0.12}_{-0.10}$ &
$2443$ & $16.19^{+0.03}_{-0.03}$ & $<0.33$ & $0.45^{+0.11}_{-0.10}$ \\
$[0.3,0.35]$ &
$4024$ & $10.44^{+0.03}_{-0.03}$ & $0.30^{+0.21}_{-0.15}$ & $0.40^{+0.11}_{-0.10}$ &
$5055$ & $12.55^{+0.03}_{-0.03}$ & $<0.37$ & $0.43^{+0.12}_{-0.10}$ &
$2958$ & $16.38^{+0.03}_{-0.03}$ & $<0.32$ & $0.45^{+0.11}_{-0.09}$ \\
$[0.35,0.4]$ &
$4379$ & $10.52^{+0.03}_{-0.03}$ & $0.35^{+0.25}_{-0.16}$ & $0.42^{+0.11}_{-0.09}$ &
$5402$ & $12.75^{+0.02}_{-0.02}$ & $<0.35$ & $0.43^{+0.11}_{-0.10}$ &
$3208$ & $16.45^{+0.03}_{-0.03}$ & $<0.34$ & $0.42^{+0.12}_{-0.09}$ \\
$[0.4,0.45]$ &
$4925$ & $10.31^{+0.03}_{-0.03}$ & $0.33^{+0.21}_{-0.17}$ & $0.41^{+0.11}_{-0.11}$ &
$5819$ & $12.45^{+0.03}_{-0.03}$ & $<0.42$ & $0.43^{+0.11}_{-0.11}$ &
$3289$ & $16.09^{+0.03}_{-0.03}$ & $<0.37$ & $0.44^{+0.11}_{-0.10}$ \\
$[0.45,0.5]$ &
$3982$ & $9.77^{+0.03}_{-0.03}$ & $0.35^{+0.21}_{-0.16}$ & $0.38^{+0.12}_{-0.09}$ &
$4813$ & $11.50^{+0.03}_{-0.03}$ & $<0.43$ & $0.41^{+0.11}_{-0.10}$ &
$2660$ & $14.69^{+0.03}_{-0.03}$ & $<0.35$ & $0.43^{+0.12}_{-0.10}$ \\
$[0.5,0.55]$ &
$2838$ & $9.25^{+0.03}_{-0.03}$ & $0.36^{+0.25}_{-0.18}$ & $0.37^{+0.13}_{-0.11}$ &
$3631$ & $10.33^{+0.03}_{-0.03}$ & $0.36^{+0.22}_{-0.16}$ & $0.42^{+0.11}_{-0.09}$ &
$2009$ & $13.02^{+0.03}_{-0.03}$ & $<0.39$ & $0.44^{+0.11}_{-0.10}$ \\
$[0.55,0.6]$ &
$1971$ & $9.06^{+0.04}_{-0.04}$ & $0.42^{+0.24}_{-0.23}$ & $0.39^{+0.11}_{-0.11}$ &
$2614$ & $10.36^{+0.05}_{-0.05}$ & $0.40^{+0.21}_{-0.18}$ & $0.41^{+0.14}_{-0.10}$ &
$1538$ & $12.38^{+0.04}_{-0.04}$ & $<0.35$ & $0.45^{+0.11}_{-0.10}$ \\
$[0.6,0.65]$ &
$1046$ & $8.85^{+0.05}_{-0.04}$ & $0.38^{+0.22}_{-0.18}$ & $0.37^{+0.12}_{-0.10}$ &
$1586$ & $9.50^{+0.05}_{-0.04}$ & $0.35^{+0.22}_{-0.17}$ & $0.38^{+0.12}_{-0.09}$ &
$977$ & $11.32^{+0.05}_{-0.05}$ & $<0.38$ & $0.42^{+0.12}_{-0.10}$ \\
\hline
\end{tabular}}
\end{minipage}
\end{table*}}}

{\renewcommand{\tabcolsep}{1.5mm}
{\renewcommand{\arraystretch}{2.}
\begin{table*}
\begin{minipage}{\textwidth}
\caption{Clusters: results from fitting Eq.~(\ref{eq:pNV}) with data from the cluster catalog from \citep{WenHan2012}.
Data and fits are divided in redshift bins (first column) and cell radius bin (as indicated by $R$ intervals). $N_{cl}$ is not a fitting parameter, but the number of clusters in each redshift and radius bin. All the fitting parameters are described in the text.}\label{tab:results_clusters}
\centering
\resizebox*{\textwidth}{!}{
\begin{tabular}{cccccccccccccc}
\hline
 & & \multicolumn{3}{c}{$R=10$ Mpc} & \multicolumn{3}{c}{$R=20$ Mpc} & \multicolumn{3}{c}{$R=30$ Mpc} & \multicolumn{3}{c}{$R=40$ Mpc} \\
            \cmidrule(lr){3-5} \cmidrule(lr){6-8} \cmidrule(lr){9-11} \cmidrule(lr){12-14}
$z$  & $N_{cl}$ & $\bar{N}$ & $b$ & $\tilde{\lambda}$
                & $\bar{N}$ & $b$ & $\tilde{\lambda}$
                & $\bar{N}$ & $b$ & $\tilde{\lambda}$
                & $\bar{N}$ & $b$ & $\tilde{\lambda}$ \\
\hline
\multicolumn{14}{c}{$\chi^{2}_{ls}$} \\
$[0.05,0.281]$ &
$37686$ & $0.071^{+0.012}_{-0.009}$ & $<0.20$ & $<636$ &
$0.49^{+0.04}_{-0.03}$ & $0.19^{+0.06}_{-0.05}$ & $<544$ &
$1.71^{+0.09}_{-0.08}$ & $0.37^{+0.06}_{-0.05}$ & $<1855$ &
$4.18^{+0.23}_{-0.21}$ & $0.49^{+0.31}_{-0.30}$ & $1854^{+684}_{-683}$\\
$[0.281,0.361]$ &
$24208$ & $0.051^{+0.015}_{-0.009}$ & $<0.29$ & $<494$ &
$0.33^{+0.03}_{-0.03}$ & $<0.16$ & $<215$ &
$1.11^{+0.06}_{-0.05}$ & $0.25^{+0.05}_{-0.04}$ & $<37$ &
$2.73^{+0.14}_{-0.12}$ & $0.43^{+0.06}_{-0.06}$ & $<3146$ \\
$[0.361,0.42]$ &
$20634$ & $0.041^{+0.012}_{-0.007}$ & $<0.27$ & $<1722$ &
$0.26^{+0.03}_{-0.02}$ & $<0.10$ & $<1.68$ &
$0.91^{+0.05}_{-0.04}$ & $0.23^{+0.05}_{-0.05}$ & $<473$ &
$2.20^{+0.11}_{-0.10}$ & $0.38^{+0.06}_{-0.05}$ & $<916$ \\
\multicolumn{14}{c}{$\chi^{2}_{jk}$} \\
$[0.05,0.281]$ &
$37686$ & $0.0623^{+0.0002}_{-0.0001}$ & $0.042^{+0.010}_{-0.008}$ & $<49$ &
$0.492^{+0.002}_{-0.002}$ & $0.18^{+0.03}_{-0.03}$ & $<7538$ &
$1.635^{+0.009}_{-0.008}$ & $<0.31$ & $<3.18$ &
$3.79^{+0.02}_{-0.02}$ & $0.45^{+0.05}_{-0.05}$ & $<12310$ \\
$[0.281,0.361]$ &
$24208$ & $0.0404^{+0.0001}_{-0.0001}$ & $0.025^{+0.005}_{-0.005}$ & $<68$ &
$0.322^{+0.001}_{-0.001}$ & $0.11^{+0.02}_{-0.02}$ & $<301$ &
$1.073^{+0.006}_{-0.005}$ & $0.24^{+0.04}_{-0.04}$ & $<1315$ &
$2.55^{+0.02}_{-0.02}$ & $0.35^{+0.05}_{-0.05}$ & $<21731$ \\
$[0.361,0.42]$ &
$20634$ & $0.03376^{+0.00009}_{-0.00010}$ & $<0.20$ & $<0.65$ &
$0.2664^{+0.0007}_{-0.0011}$ & $<0.071$ & $<1.33$ &
$0.885^{+0.004}_{-0.004}$ & $0.20^{+0.04}_{-0.03}$ & $<6130$ &
$2.07^{+0.01}_{-0.01}$ & $0.30^{+0.05}_{-0.04}$ & $<113$ \\
\hline
\end{tabular}}
\end{minipage}
\end{table*}}}

\begin{figure*}
\includegraphics[width=\textwidth]{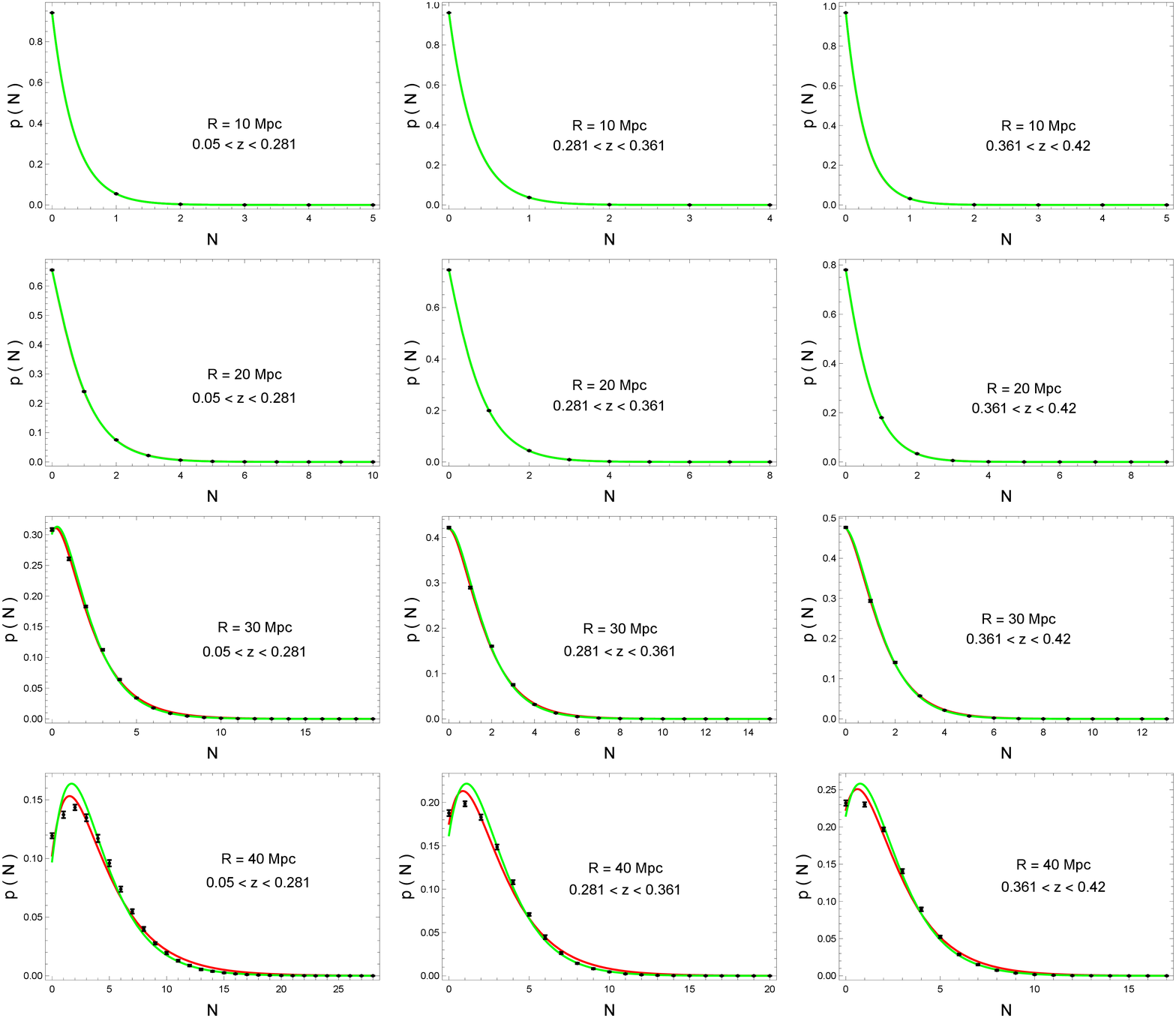}
\caption{Comparison between data and theoretical expectation Eq.~(\ref{eq:pNV}) for cell size bin $1.0 < R_{1} < 1.1$ $Mpc$. Black points: data; black bars: jackknife-like observational errors. Solid red line: best fit from minimization of $\chi_{ls}^{2}$; solid green line: best fit from minimization of $\chi_{jk}^{2}$.}\label{fig:CL}
\end{figure*}

\section{Discussion and Conclusions}

Nonlocality effects  can occur at large distances as the consequence of modified   gravitational potential. Specifically, nonlocal gravity, besides the  effects at cosmological scale as a possible engine for accelerated expansion \citep{50acd,Nunes}, is relevant also for large scale formation. In particular, it is capable of triggering the clustering of galaxies and clusters of galaxies. 

In view of this statement,  we have analyzed the effects of nonlocal gravitational potential on the structure formation comparing the model with observational data.  This has been done by treating both galaxies and clusters of galaxies (in two distinct analysis) as points in a statistical mechanical system, and then analyzing their clustering using a gravitational partition function. This  function has been modified according to the nonlocal gravitational potential and its influence on the properties of the gravitational clustering has been scrutinized. Finally, the model has also been compared with  observational data, and we have demonstrated that a logarithmic correction to standard gravity seems to be consistent with the observations, at least with those concerning the galactic scale, i.e. within volumes corresponding to clusters of galaxies. Thus, it is possible to conclude that the nonlocal gravitational effects might be revealed at such scales. On the other hand, we must stress that the clustering of larger volumes of space, in which the points are the clusters of galaxies, does not show any evidence in favor of a such modification to gravity so that any inference should be suspended until further (maybe more precise) probes are checked.

For a more general discussion, it has be noted that other modifications of gravitational potential,  explaining  dark matter or dark energy effects, have been studied considering  modifications of the gravitational partition function. For example, it is well-known that  $f(R)$ gravity modifies the  large distance behavior of gravitational potential \citep{Curvature,Report,gravity,Oikonomou}. Thus, the clustering of galaxies, interacting through effective  potentials derived from  $f(R)$ gravity, has also been studied using a modified gravitational partition function, and it has been observed that also these models are consistent with observations for  large samples of galaxies and clusters \citep{Betty,Cap2}, although the corresponding statistical inferenec is a bit weaker that the nonlocal scenario we have considered here. It is worth  noticing that the clustering of galaxies in $f(R)$ gravity is also consistent with the constraints  coming from  the \textit{Planck} data \citep{2az}. 

The MOND  gravitational partition function has  also been used to study the thermodynamics and the clustering properties of  systems of galaxies \citep{3,4}. 


Furthermore, the gravitational phase transition can also be analyzed using the gravitational partition function for  systems of galaxies \citep{2b}. In fact, a first order phase transition occurs  due to the clustering of galaxies  from a homogeneous phase. In this framework, it is  possible to take into account the Yang-Lee theory for systems of galaxies, and use the complex fugacity  to analyze the  phase transition  \citep{4b}. A forthcoming step will be to extend the Yang-Lee theory for nonlocal gravitational potential, and analyze the phase transition for the related  gravitational partition function. 

In particular, it has to be noted that the  clustering  of galaxies has also been investigated using the cosmic energy equation \citep{Ahmad}.  As the cosmic energy equation is  derived by approximating galaxies as points in a statistical mechanical system, a softening parameter can  modify the cosmic energy equation \citep{ce}. It has been demonstrated that a large scale modification of the gravitational potential also modifies the  cosmic energy equation \citep{Hameeda}. 

\section*{Acknowledgements}
SC acknowledgs the support of the  Istituto Nazionale di Fisica Nucleare (INFN), sezione di Napoli, iniziative specifiche QGSKY and MOONLIGHT-2.

\bibliographystyle{apsrev4-1}
\bibliography{biblio.bib}

\begin{thebibliography}{89}%
\makeatletter
\providecommand \@ifxundefined [1]{%
 \@ifx{#1\undefined}
}%
\providecommand \@ifnum [1]{%
 \ifnum #1\expandafter \@firstoftwo
 \else \expandafter \@secondoftwo
 \fi
}%
\providecommand \@ifx [1]{%
 \ifx #1\expandafter \@firstoftwo
 \else \expandafter \@secondoftwo
 \fi
}%
\providecommand \natexlab [1]{#1}%
\providecommand \enquote  [1]{``#1''}%
\providecommand \bibnamefont  [1]{#1}%
\providecommand \bibfnamefont [1]{#1}%
\providecommand \citenamefont [1]{#1}%
\providecommand \href@noop [0]{\@secondoftwo}%
\providecommand \href [0]{\begingroup \@sanitize@url \@href}%
\providecommand \@href[1]{\@@startlink{#1}\@@href}%
\providecommand \@@href[1]{\endgroup#1\@@endlink}%
\providecommand \@sanitize@url [0]{\catcode `\\12\catcode `\$12\catcode
  `\&12\catcode `\#12\catcode `\^12\catcode `\_12\catcode `\%12\relax}%
\providecommand \@@startlink[1]{}%
\providecommand \@@endlink[0]{}%
\providecommand \url  [0]{\begingroup\@sanitize@url \@url }%
\providecommand \@url [1]{\endgroup\@href {#1}{\urlprefix }}%
\providecommand \urlprefix  [0]{URL }%
\providecommand \Eprint [0]{\href }%
\providecommand \doibase [0]{http://dx.doi.org/}%
\providecommand \selectlanguage [0]{\@gobble}%
\providecommand \bibinfo  [0]{\@secondoftwo}%
\providecommand \bibfield  [0]{\@secondoftwo}%
\providecommand \translation [1]{[#1]}%
\providecommand \BibitemOpen [0]{}%
\providecommand \bibitemStop [0]{}%
\providecommand \bibitemNoStop [0]{.\EOS\space}%
\providecommand \EOS [0]{\spacefactor3000\relax}%
\providecommand \BibitemShut  [1]{\csname bibitem#1\endcsname}%
\let\auto@bib@innerbib\@empty
\bibitem [{\citenamefont {Frenk}\ \emph {et~al.}(1999)\citenamefont {Frenk}
  \emph {et~al.}}]{a1}%
  \BibitemOpen
  \bibfield  {author} {\bibinfo {author} {\bibfnamefont {C.~S.}\ \bibnamefont
  {Frenk}} \emph {et~al.},\ }\href {\doibase 10.1086/307908} {\bibfield
  {journal} {\bibinfo  {journal} {Astrophys. J.}\ }\textbf {\bibinfo {volume}
  {525}},\ \bibinfo {pages} {554} (\bibinfo {year} {1999})},\ \Eprint
  {http://arxiv.org/abs/astro-ph/9906160} {arXiv:astro-ph/9906160} \BibitemShut
  {NoStop}%
\bibitem [{\citenamefont {{Barnes}}\ \emph {et~al.}(2017)\citenamefont
  {{Barnes}}, \citenamefont {{Kay}}, \citenamefont {{Bah{\'e}}}, \citenamefont
  {{Dalla Vecchia}}, \citenamefont {{McCarthy}}, \citenamefont {{Schaye}},
  \citenamefont {{Bower}}, \citenamefont {{Jenkins}}, \citenamefont {{Thomas}},
  \citenamefont {{Schaller}}, \citenamefont {{Crain}}, \citenamefont
  {{Theuns}},\ and\ \citenamefont {{White}}}]{a2}%
  \BibitemOpen
  \bibfield  {author} {\bibinfo {author} {\bibfnamefont {D.~J.}\ \bibnamefont
  {{Barnes}}}, \bibinfo {author} {\bibfnamefont {S.~T.}\ \bibnamefont {{Kay}}},
  \bibinfo {author} {\bibfnamefont {Y.~M.}\ \bibnamefont {{Bah{\'e}}}},
  \bibinfo {author} {\bibfnamefont {C.}~\bibnamefont {{Dalla Vecchia}}},
  \bibinfo {author} {\bibfnamefont {I.~G.}\ \bibnamefont {{McCarthy}}},
  \bibinfo {author} {\bibfnamefont {J.}~\bibnamefont {{Schaye}}}, \bibinfo
  {author} {\bibfnamefont {R.~G.}\ \bibnamefont {{Bower}}}, \bibinfo {author}
  {\bibfnamefont {A.}~\bibnamefont {{Jenkins}}}, \bibinfo {author}
  {\bibfnamefont {P.~A.}\ \bibnamefont {{Thomas}}}, \bibinfo {author}
  {\bibfnamefont {M.}~\bibnamefont {{Schaller}}}, \bibinfo {author}
  {\bibfnamefont {R.~A.}\ \bibnamefont {{Crain}}}, \bibinfo {author}
  {\bibfnamefont {T.}~\bibnamefont {{Theuns}}}, \ and\ \bibinfo {author}
  {\bibfnamefont {S.~D.~M.}\ \bibnamefont {{White}}},\ }\href {\doibase
  10.1093/mnras/stx1647} {\bibfield  {journal} {\bibinfo  {journal} {Mon. Not.
  Roy. Astron. Soc.}\ }\textbf {\bibinfo {volume} {471}},\ \bibinfo {pages}
  {1088} (\bibinfo {year} {2017})},\ \Eprint {http://arxiv.org/abs/1703.10907}
  {arXiv:1703.10907 [astro-ph.GA]} \BibitemShut {NoStop}%
\bibitem [{\citenamefont {Ponman}\ \emph {et~al.}(2003)\citenamefont {Ponman},
  \citenamefont {Sanderson},\ and\ \citenamefont {Finoguenov}}]{a3}%
  \BibitemOpen
  \bibfield  {author} {\bibinfo {author} {\bibfnamefont {T.~J.}\ \bibnamefont
  {Ponman}}, \bibinfo {author} {\bibfnamefont {A.~J.~R.}\ \bibnamefont
  {Sanderson}}, \ and\ \bibinfo {author} {\bibfnamefont {A.}~\bibnamefont
  {Finoguenov}},\ }\href {\doibase 10.1046/j.1365-8711.2003.06677.x} {\bibfield
   {journal} {\bibinfo  {journal} {Mon. Not. Roy. Astron. Soc.}\ }\textbf
  {\bibinfo {volume} {343}},\ \bibinfo {pages} {331} (\bibinfo {year}
  {2003})},\ \Eprint {http://arxiv.org/abs/astro-ph/0304048}
  {arXiv:astro-ph/0304048} \BibitemShut {NoStop}%
\bibitem [{\citenamefont {Voit}(2005)}]{a4}%
  \BibitemOpen
  \bibfield  {author} {\bibinfo {author} {\bibfnamefont {G.~M.}\ \bibnamefont
  {Voit}},\ }\href {\doibase 10.1103/RevModPhys.77.207} {\bibfield  {journal}
  {\bibinfo  {journal} {Rev. Mod. Phys.}\ }\textbf {\bibinfo {volume} {77}},\
  \bibinfo {pages} {207} (\bibinfo {year} {2005})},\ \Eprint
  {http://arxiv.org/abs/astro-ph/0410173} {arXiv:astro-ph/0410173} \BibitemShut
  {NoStop}%
\bibitem [{\citenamefont {{Saslaw}}\ and\ \citenamefont {{Yang}}(2009)}]{cos1}%
  \BibitemOpen
  \bibfield  {author} {\bibinfo {author} {\bibfnamefont {W.~C.}\ \bibnamefont
  {{Saslaw}}}\ and\ \bibinfo {author} {\bibfnamefont {A.}~\bibnamefont
  {{Yang}}},\ }\href@noop {} {\bibfield  {journal} {\bibinfo  {journal} {arXiv
  e-prints}\ ,\ \bibinfo {eid} {arXiv:0902.0747}} (\bibinfo {year} {2009})},\
  \Eprint {http://arxiv.org/abs/0902.0747} {arXiv:0902.0747 [astro-ph.CO]}
  \BibitemShut {NoStop}%
\bibitem [{\citenamefont {{Saslaw}}\ and\ \citenamefont {{Fang}}(1996)}]{cos2}%
  \BibitemOpen
  \bibfield  {author} {\bibinfo {author} {\bibfnamefont {W.~C.}\ \bibnamefont
  {{Saslaw}}}\ and\ \bibinfo {author} {\bibfnamefont {F.}~\bibnamefont
  {{Fang}}},\ }\href {\doibase 10.1086/176949} {\bibfield  {journal} {\bibinfo
  {journal} {Astrophys. J.}\ }\textbf {\bibinfo {volume} {460}},\ \bibinfo
  {pages} {16} (\bibinfo {year} {1996})}\BibitemShut {NoStop}%
\bibitem [{\citenamefont {{Saslaw}}\ and\ \citenamefont
  {{Sheth}}(1993)}]{cos3}%
  \BibitemOpen
  \bibfield  {author} {\bibinfo {author} {\bibfnamefont {W.~C.}\ \bibnamefont
  {{Saslaw}}}\ and\ \bibinfo {author} {\bibfnamefont {R.~K.}\ \bibnamefont
  {{Sheth}}},\ }\href {\doibase 10.1086/172682} {\bibfield  {journal} {\bibinfo
   {journal} {Astrophys. J.}\ }\textbf {\bibinfo {volume} {409}},\ \bibinfo
  {pages} {504} (\bibinfo {year} {1993})}\BibitemShut {NoStop}%
\bibitem [{\citenamefont {Rahmani}\ \emph {et~al.}(2009)\citenamefont
  {Rahmani}, \citenamefont {Saslaw},\ and\ \citenamefont {Tavasoli}}]{cos4}%
  \BibitemOpen
  \bibfield  {author} {\bibinfo {author} {\bibfnamefont {H.}~\bibnamefont
  {Rahmani}}, \bibinfo {author} {\bibfnamefont {W.~C.}\ \bibnamefont {Saslaw}},
  \ and\ \bibinfo {author} {\bibfnamefont {S.}~\bibnamefont {Tavasoli}},\
  }\href {\doibase 10.1088/0004-637X/695/2/1121} {\bibfield  {journal}
  {\bibinfo  {journal} {Astrophys. J.}\ }\textbf {\bibinfo {volume} {695}},\
  \bibinfo {pages} {1121} (\bibinfo {year} {2009})},\ \Eprint
  {http://arxiv.org/abs/0902.1103} {arXiv:0902.1103 [astro-ph.CO]} \BibitemShut
  {NoStop}%
\bibitem [{\citenamefont {{Itoh}}\ \emph {et~al.}(1993)\citenamefont {{Itoh}},
  \citenamefont {{Inagaki}},\ and\ \citenamefont {{Saslaw}}}]{cosa1}%
  \BibitemOpen
  \bibfield  {author} {\bibinfo {author} {\bibfnamefont {M.}~\bibnamefont
  {{Itoh}}}, \bibinfo {author} {\bibfnamefont {S.}~\bibnamefont {{Inagaki}}}, \
  and\ \bibinfo {author} {\bibfnamefont {W.~C.}\ \bibnamefont {{Saslaw}}},\
  }\href {\doibase 10.1086/172219} {\bibfield  {journal} {\bibinfo  {journal}
  {Astrophys. J.}\ }\textbf {\bibinfo {volume} {403}},\ \bibinfo {pages} {476}
  (\bibinfo {year} {1993})}\BibitemShut {NoStop}%
\bibitem [{\citenamefont {{Saslaw}}\ \emph {et~al.}(1990)\citenamefont
  {{Saslaw}}, \citenamefont {{Chitre}}, \citenamefont {{Itoh}},\ and\
  \citenamefont {{Inagaki}}}]{sas2}%
  \BibitemOpen
  \bibfield  {author} {\bibinfo {author} {\bibfnamefont {W.~C.}\ \bibnamefont
  {{Saslaw}}}, \bibinfo {author} {\bibfnamefont {S.~M.}\ \bibnamefont
  {{Chitre}}}, \bibinfo {author} {\bibfnamefont {M.}~\bibnamefont {{Itoh}}}, \
  and\ \bibinfo {author} {\bibfnamefont {S.}~\bibnamefont {{Inagaki}}},\ }\href
  {\doibase 10.1086/169496} {\bibfield  {journal} {\bibinfo  {journal}
  {Astrophys. J.}\ }\textbf {\bibinfo {volume} {365}},\ \bibinfo {pages} {419}
  (\bibinfo {year} {1990})}\BibitemShut {NoStop}%
\bibitem [{\citenamefont {{Saslaw}}(1986)}]{sas}%
  \BibitemOpen
  \bibfield  {author} {\bibinfo {author} {\bibfnamefont {W.~C.}\ \bibnamefont
  {{Saslaw}}},\ }\href {\doibase 10.1086/164142} {\bibfield  {journal}
  {\bibinfo  {journal} {Astrophys. J.}\ }\textbf {\bibinfo {volume} {304}},\
  \bibinfo {pages} {11} (\bibinfo {year} {1986})}\BibitemShut {NoStop}%
\bibitem [{\citenamefont {{Saslaw}}\ and\ \citenamefont
  {{Ahmad}}(2010)}]{sin12}%
  \BibitemOpen
  \bibfield  {author} {\bibinfo {author} {\bibfnamefont {W.~C.}\ \bibnamefont
  {{Saslaw}}}\ and\ \bibinfo {author} {\bibfnamefont {F.}~\bibnamefont
  {{Ahmad}}},\ }\href {\doibase 10.1088/0004-637X/720/2/1246} {\bibfield
  {journal} {\bibinfo  {journal} {Astrophys. J.}\ }\textbf {\bibinfo {volume}
  {720}},\ \bibinfo {pages} {1246} (\bibinfo {year} {2010})}\BibitemShut
  {NoStop}%
\bibitem [{\citenamefont {Upadhyay}\ \emph {et~al.}(2018)\citenamefont
  {Upadhyay}, \citenamefont {Pourhassan},\ and\ \citenamefont
  {Capozziello}}]{3}%
  \BibitemOpen
  \bibfield  {author} {\bibinfo {author} {\bibfnamefont {S.}~\bibnamefont
  {Upadhyay}}, \bibinfo {author} {\bibfnamefont {B.}~\bibnamefont
  {Pourhassan}}, \ and\ \bibinfo {author} {\bibfnamefont {S.}~\bibnamefont
  {Capozziello}},\ }\href {\doibase 10.1142/S0218271819500275} {\bibfield
  {journal} {\bibinfo  {journal} {Int. J. Mod. Phys. D}\ }\textbf {\bibinfo
  {volume} {28}},\ \bibinfo {pages} {1950027} (\bibinfo {year} {2018})},\
  \Eprint {http://arxiv.org/abs/1809.03579} {arXiv:1809.03579 [gr-qc]}
  \BibitemShut {NoStop}%
\bibitem [{\citenamefont {{Khan}}\ and\ \citenamefont {{Malik}}(2012)}]{2b}%
  \BibitemOpen
  \bibfield  {author} {\bibinfo {author} {\bibfnamefont {M.~S.}\ \bibnamefont
  {{Khan}}}\ and\ \bibinfo {author} {\bibfnamefont {M.~A.}\ \bibnamefont
  {{Malik}}},\ }\href {\doibase 10.1111/j.1365-2966.2012.20493.x} {\bibfield
  {journal} {\bibinfo  {journal} {Mon. Not. Roy. Astron. Soc.}\ }\textbf
  {\bibinfo {volume} {421}},\ \bibinfo {pages} {2629} (\bibinfo {year}
  {2012})}\BibitemShut {NoStop}%
\bibitem [{\citenamefont {{Ahmad}}\ and\ \citenamefont
  {{Hameeda}}(2010)}]{ahm10}%
  \BibitemOpen
  \bibfield  {author} {\bibinfo {author} {\bibfnamefont {F.}~\bibnamefont
  {{Ahmad}}}\ and\ \bibinfo {author} {\bibnamefont {{Hameeda}}},\ }\href
  {\doibase 10.1007/s10509-010-0416-9} {\bibfield  {journal} {\bibinfo
  {journal} {Astrophysics and Space Science}\ }\textbf {\bibinfo {volume}
  {330}},\ \bibinfo {pages} {227} (\bibinfo {year} {2010})}\BibitemShut
  {NoStop}%
\bibitem [{\citenamefont {Hameeda}\ \emph {et~al.}(2021)\citenamefont
  {Hameeda}, \citenamefont {Pourhassan}, \citenamefont {Rocca},\ and\
  \citenamefont {Brzo}}]{reference_epjc}%
  \BibitemOpen
  \bibfield  {author} {\bibinfo {author} {\bibfnamefont {M.}~\bibnamefont
  {Hameeda}}, \bibinfo {author} {\bibfnamefont {B.}~\bibnamefont {Pourhassan}},
  \bibinfo {author} {\bibfnamefont {M.~C.}\ \bibnamefont {Rocca}}, \ and\
  \bibinfo {author} {\bibfnamefont {A.~B.}\ \bibnamefont {Brzo}},\ }\href
  {\doibase 10.1140/epjc/s10052-021-08940-0} {\bibfield  {journal} {\bibinfo
  {journal} {Eur. Phys. J. C}\ }\textbf {\bibinfo {volume} {81}},\ \bibinfo
  {pages} {146} (\bibinfo {year} {2021})}\BibitemShut {NoStop}%
\bibitem [{\citenamefont {{Saslaw}}\ and\ \citenamefont
  {{Hamilton}}(1984)}]{sas84}%
  \BibitemOpen
  \bibfield  {author} {\bibinfo {author} {\bibfnamefont {W.~C.}\ \bibnamefont
  {{Saslaw}}}\ and\ \bibinfo {author} {\bibfnamefont {A.~J.~S.}\ \bibnamefont
  {{Hamilton}}},\ }\href {\doibase 10.1086/161589} {\bibfield  {journal}
  {\bibinfo  {journal} {Astrophys. J.}\ }\textbf {\bibinfo {volume} {276}},\
  \bibinfo {pages} {13} (\bibinfo {year} {1984})}\BibitemShut {NoStop}%
\bibitem [{\citenamefont {{Ahmad}}\ \emph {et~al.}(2002)\citenamefont
  {{Ahmad}}, \citenamefont {{Saslaw}},\ and\ \citenamefont {{Bhat}}}]{ahm02}%
  \BibitemOpen
  \bibfield  {author} {\bibinfo {author} {\bibfnamefont {F.}~\bibnamefont
  {{Ahmad}}}, \bibinfo {author} {\bibfnamefont {W.~C.}\ \bibnamefont
  {{Saslaw}}}, \ and\ \bibinfo {author} {\bibfnamefont {N.~I.}\ \bibnamefont
  {{Bhat}}},\ }\href {\doibase 10.1086/340095} {\bibfield  {journal} {\bibinfo
  {journal} {Astrophys. J.}\ }\textbf {\bibinfo {volume} {571}},\ \bibinfo
  {pages} {576} (\bibinfo {year} {2002})}\BibitemShut {NoStop}%
\bibitem [{\citenamefont {{Sivakoff}}\ and\ \citenamefont
  {{Saslaw}}(2005)}]{10}%
  \BibitemOpen
  \bibfield  {author} {\bibinfo {author} {\bibfnamefont {G.~R.}\ \bibnamefont
  {{Sivakoff}}}\ and\ \bibinfo {author} {\bibfnamefont {W.~C.}\ \bibnamefont
  {{Saslaw}}},\ }\href {\doibase 10.1086/430082} {\bibfield  {journal}
  {\bibinfo  {journal} {Astrophys. J.}\ }\textbf {\bibinfo {volume} {626}},\
  \bibinfo {pages} {795} (\bibinfo {year} {2005})},\ \Eprint
  {http://arxiv.org/abs/astro-ph/0503509} {arXiv:astro-ph/0503509 [astro-ph]}
  \BibitemShut {NoStop}%
\bibitem [{\citenamefont {{Rahmani}}\ \emph {et~al.}(2009)\citenamefont
  {{Rahmani}}, \citenamefont {{Saslaw}},\ and\ \citenamefont
  {{Tavasoli}}}]{20}%
  \BibitemOpen
  \bibfield  {author} {\bibinfo {author} {\bibfnamefont {H.}~\bibnamefont
  {{Rahmani}}}, \bibinfo {author} {\bibfnamefont {W.~C.}\ \bibnamefont
  {{Saslaw}}}, \ and\ \bibinfo {author} {\bibfnamefont {S.}~\bibnamefont
  {{Tavasoli}}},\ }\href {\doibase 10.1088/0004-637X/695/2/1121} {\bibfield
  {journal} {\bibinfo  {journal} {Astrophys. J.}\ }\textbf {\bibinfo {volume}
  {695}},\ \bibinfo {pages} {1121} (\bibinfo {year} {2009})},\ \Eprint
  {http://arxiv.org/abs/0902.1103} {arXiv:0902.1103 [astro-ph.CO]} \BibitemShut
  {NoStop}%
\bibitem [{\citenamefont {{Ahmad}}\ \emph {et~al.}(2006)\citenamefont
  {{Ahmad}}, \citenamefont {{Saslaw}},\ and\ \citenamefont {{Malik}}}]{ahm06}%
  \BibitemOpen
  \bibfield  {author} {\bibinfo {author} {\bibfnamefont {F.}~\bibnamefont
  {{Ahmad}}}, \bibinfo {author} {\bibfnamefont {W.~C.}\ \bibnamefont
  {{Saslaw}}}, \ and\ \bibinfo {author} {\bibfnamefont {M.~A.}\ \bibnamefont
  {{Malik}}},\ }\href {\doibase 10.1086/504396} {\bibfield  {journal} {\bibinfo
   {journal} {Astrophys. J.}\ }\textbf {\bibinfo {volume} {645}},\ \bibinfo
  {pages} {940} (\bibinfo {year} {2006})}\BibitemShut {NoStop}%
\bibitem [{\citenamefont {Riess}\ \emph {et~al.}(1998)\citenamefont {Riess}
  \emph {et~al.}}]{1ab}%
  \BibitemOpen
  \bibfield  {author} {\bibinfo {author} {\bibfnamefont {A.~G.}\ \bibnamefont
  {Riess}} \emph {et~al.} (\bibinfo {collaboration} {Supernova Search Team}),\
  }\href {\doibase 10.1086/300499} {\bibfield  {journal} {\bibinfo  {journal}
  {Astron. J.}\ }\textbf {\bibinfo {volume} {116}},\ \bibinfo {pages} {1009}
  (\bibinfo {year} {1998})},\ \Eprint {http://arxiv.org/abs/astro-ph/9805201}
  {arXiv:astro-ph/9805201} \BibitemShut {NoStop}%
\bibitem [{\citenamefont {Perlmutter}\ \emph {et~al.}(1999)\citenamefont
  {Perlmutter} \emph {et~al.}}]{6ab}%
  \BibitemOpen
  \bibfield  {author} {\bibinfo {author} {\bibfnamefont {S.}~\bibnamefont
  {Perlmutter}} \emph {et~al.} (\bibinfo {collaboration} {Supernova Cosmology
  Project}),\ }\href {\doibase 10.1086/307221} {\bibfield  {journal} {\bibinfo
  {journal} {Astrophys. J.}\ }\textbf {\bibinfo {volume} {517}},\ \bibinfo
  {pages} {565} (\bibinfo {year} {1999})},\ \Eprint
  {http://arxiv.org/abs/astro-ph/9812133} {arXiv:astro-ph/9812133} \BibitemShut
  {NoStop}%
\bibitem [{\citenamefont {Aghanim}\ \emph {et~al.}(2020)\citenamefont {Aghanim}
  \emph {et~al.}}]{Aghanim:2018eyx}%
  \BibitemOpen
  \bibfield  {author} {\bibinfo {author} {\bibfnamefont {N.}~\bibnamefont
  {Aghanim}} \emph {et~al.} (\bibinfo {collaboration} {Planck}),\ }\href
  {\doibase 10.1051/0004-6361/201833910} {\bibfield  {journal} {\bibinfo
  {journal} {Astron. Astrophys.}\ }\textbf {\bibinfo {volume} {641}},\ \bibinfo
  {pages} {A6} (\bibinfo {year} {2020})},\ \Eprint
  {http://arxiv.org/abs/1807.06209} {arXiv:1807.06209 [astro-ph.CO]}
  \BibitemShut {NoStop}%
\bibitem [{\citenamefont {Gurzadyan}\ \emph {et~al.}(2020)\citenamefont
  {Gurzadyan}, \citenamefont {Kocharyan},\ and\ \citenamefont
  {Stepanian}}]{cosd1}%
  \BibitemOpen
  \bibfield  {author} {\bibinfo {author} {\bibfnamefont {V.~G.}\ \bibnamefont
  {Gurzadyan}}, \bibinfo {author} {\bibfnamefont {A.~A.}\ \bibnamefont
  {Kocharyan}}, \ and\ \bibinfo {author} {\bibfnamefont {A.}~\bibnamefont
  {Stepanian}},\ }\href {\doibase 10.1140/epjc/s10052-019-7597-3} {\bibfield
  {journal} {\bibinfo  {journal} {Eur. Phys. J. C}\ }\textbf {\bibinfo {volume}
  {80}},\ \bibinfo {pages} {24} (\bibinfo {year} {2020})},\ \Eprint
  {http://arxiv.org/abs/2001.02634} {arXiv:2001.02634 [astro-ph.CO]}
  \BibitemShut {NoStop}%
\bibitem [{\citenamefont {Gurzadyan}\ and\ \citenamefont
  {Stepanian}(2018)}]{cosd2}%
  \BibitemOpen
  \bibfield  {author} {\bibinfo {author} {\bibfnamefont {V.~G.}\ \bibnamefont
  {Gurzadyan}}\ and\ \bibinfo {author} {\bibfnamefont {A.}~\bibnamefont
  {Stepanian}},\ }\href {\doibase 10.1140/epjc/s10052-018-6358-z} {\bibfield
  {journal} {\bibinfo  {journal} {Eur. Phys. J. C}\ }\textbf {\bibinfo {volume}
  {78}},\ \bibinfo {pages} {869} (\bibinfo {year} {2018})},\ \Eprint
  {http://arxiv.org/abs/1810.09846} {arXiv:1810.09846 [physics.gen-ph]}
  \BibitemShut {NoStop}%
\bibitem [{\citenamefont {Gurzadyan}\ and\ \citenamefont
  {Stepanian}(2019{\natexlab{a}})}]{cosd3}%
  \BibitemOpen
  \bibfield  {author} {\bibinfo {author} {\bibfnamefont {V.~G.}\ \bibnamefont
  {Gurzadyan}}\ and\ \bibinfo {author} {\bibfnamefont {A.}~\bibnamefont
  {Stepanian}},\ }\href {\doibase 10.1140/epjc/s10052-019-6685-8} {\bibfield
  {journal} {\bibinfo  {journal} {Eur. Phys. J. C}\ }\textbf {\bibinfo {volume}
  {79}},\ \bibinfo {pages} {169} (\bibinfo {year} {2019}{\natexlab{a}})},\
  \Eprint {http://arxiv.org/abs/1902.07171} {arXiv:1902.07171 [physics.gen-ph]}
  \BibitemShut {NoStop}%
\bibitem [{\citenamefont {Gurzadyan}\ and\ \citenamefont
  {Penrose}(2013)}]{cosd4}%
  \BibitemOpen
  \bibfield  {author} {\bibinfo {author} {\bibfnamefont {V.~G.}\ \bibnamefont
  {Gurzadyan}}\ and\ \bibinfo {author} {\bibfnamefont {R.}~\bibnamefont
  {Penrose}},\ }\href {\doibase 10.1140/epjp/i2013-13022-4} {\bibfield
  {journal} {\bibinfo  {journal} {Eur. Phys. J. Plus}\ }\textbf {\bibinfo
  {volume} {128}},\ \bibinfo {pages} {22} (\bibinfo {year} {2013})},\ \Eprint
  {http://arxiv.org/abs/1302.5162} {arXiv:1302.5162 [astro-ph.CO]} \BibitemShut
  {NoStop}%
\bibitem [{\citenamefont {Gurzadyan}\ and\ \citenamefont
  {Stepanian}(2019{\natexlab{b}})}]{cosd5}%
  \BibitemOpen
  \bibfield  {author} {\bibinfo {author} {\bibfnamefont {V.~G.}\ \bibnamefont
  {Gurzadyan}}\ and\ \bibinfo {author} {\bibfnamefont {A.}~\bibnamefont
  {Stepanian}},\ }\href {\doibase 10.1140/epjc/s10052-019-7081-0} {\bibfield
  {journal} {\bibinfo  {journal} {Eur. Phys. J. C}\ }\textbf {\bibinfo {volume}
  {79}},\ \bibinfo {pages} {568} (\bibinfo {year} {2019}{\natexlab{b}})},\
  \Eprint {http://arxiv.org/abs/1905.03442} {arXiv:1905.03442 [astro-ph.CO]}
  \BibitemShut {NoStop}%
\bibitem [{\citenamefont {Le~Delliou}\ \emph {et~al.}(2019)\citenamefont
  {Le~Delliou}, \citenamefont {Marcondes},\ and\ \citenamefont
  {Neto}}]{cosd51}%
  \BibitemOpen
  \bibfield  {author} {\bibinfo {author} {\bibfnamefont {M.}~\bibnamefont
  {Le~Delliou}}, \bibinfo {author} {\bibfnamefont {R.~J.~F.}\ \bibnamefont
  {Marcondes}}, \ and\ \bibinfo {author} {\bibfnamefont {G.~a. B.~L.}\
  \bibnamefont {Neto}},\ }\href {\doibase 10.1093/mnras/stz2757} {\bibfield
  {journal} {\bibinfo  {journal} {Mon. Not. Roy. Astron. Soc.}\ }\textbf
  {\bibinfo {volume} {490}},\ \bibinfo {pages} {1944} (\bibinfo {year}
  {2019})},\ \Eprint {http://arxiv.org/abs/1811.10712} {arXiv:1811.10712
  [astro-ph.CO]} \BibitemShut {NoStop}%
\bibitem [{\citenamefont {Boonserm}\ \emph {et~al.}(2020)\citenamefont
  {Boonserm}, \citenamefont {Ngampitipan}, \citenamefont {Simpson},\ and\
  \citenamefont {Visser}}]{cosd6}%
  \BibitemOpen
  \bibfield  {author} {\bibinfo {author} {\bibfnamefont {P.}~\bibnamefont
  {Boonserm}}, \bibinfo {author} {\bibfnamefont {T.}~\bibnamefont
  {Ngampitipan}}, \bibinfo {author} {\bibfnamefont {A.}~\bibnamefont
  {Simpson}}, \ and\ \bibinfo {author} {\bibfnamefont {M.}~\bibnamefont
  {Visser}},\ }\href {\doibase 10.1103/PhysRevD.101.024050} {\bibfield
  {journal} {\bibinfo  {journal} {Phys. Rev. D}\ }\textbf {\bibinfo {volume}
  {101}},\ \bibinfo {pages} {024050} (\bibinfo {year} {2020})},\ \Eprint
  {http://arxiv.org/abs/1909.06755} {arXiv:1909.06755 [gr-qc]} \BibitemShut
  {NoStop}%
\bibitem [{\citenamefont {Peirani}\ and\ \citenamefont
  {Pacheco}(2008)}]{cosd71}%
  \BibitemOpen
  \bibfield  {author} {\bibinfo {author} {\bibfnamefont {S.}~\bibnamefont
  {Peirani}}\ and\ \bibinfo {author} {\bibfnamefont {J.~A. D.~F.}\ \bibnamefont
  {Pacheco}},\ }\href {\doibase 10.1051/0004-6361:200809711} {\bibfield
  {journal} {\bibinfo  {journal} {Astron. Astrophys.}\ }\textbf {\bibinfo
  {volume} {488}},\ \bibinfo {pages} {845} (\bibinfo {year} {2008})},\ \Eprint
  {http://arxiv.org/abs/0806.4245} {arXiv:0806.4245 [astro-ph]} \BibitemShut
  {NoStop}%
\bibitem [{\citenamefont {Iliev}\ and\ \citenamefont {Shapiro}(2001)}]{cosd7}%
  \BibitemOpen
  \bibfield  {author} {\bibinfo {author} {\bibfnamefont {I.~T.}\ \bibnamefont
  {Iliev}}\ and\ \bibinfo {author} {\bibfnamefont {P.~R.}\ \bibnamefont
  {Shapiro}},\ }\href {\doibase 10.1046/j.1365-8711.2001.04422.x} {\bibfield
  {journal} {\bibinfo  {journal} {Mon. Not. Roy. Astron. Soc.}\ }\textbf
  {\bibinfo {volume} {325}},\ \bibinfo {pages} {468} (\bibinfo {year}
  {2001})},\ \Eprint {http://arxiv.org/abs/astro-ph/0101067}
  {arXiv:astro-ph/0101067} \BibitemShut {NoStop}%
\bibitem [{\citenamefont {Stark}\ \emph {et~al.}(2017)\citenamefont {Stark},
  \citenamefont {Miller},\ and\ \citenamefont {Huterer}}]{cosd8}%
  \BibitemOpen
  \bibfield  {author} {\bibinfo {author} {\bibfnamefont {A.}~\bibnamefont
  {Stark}}, \bibinfo {author} {\bibfnamefont {C.~J.}\ \bibnamefont {Miller}}, \
  and\ \bibinfo {author} {\bibfnamefont {D.}~\bibnamefont {Huterer}},\ }\href
  {\doibase 10.1103/PhysRevD.96.023543} {\bibfield  {journal} {\bibinfo
  {journal} {Phys. Rev. D}\ }\textbf {\bibinfo {volume} {96}},\ \bibinfo
  {pages} {023543} (\bibinfo {year} {2017})},\ \Eprint
  {http://arxiv.org/abs/1611.06886} {arXiv:1611.06886 [astro-ph.CO]}
  \BibitemShut {NoStop}%
\bibitem [{\citenamefont {Aguena}\ and\ \citenamefont {Lima}(2018)}]{cosd9}%
  \BibitemOpen
  \bibfield  {author} {\bibinfo {author} {\bibfnamefont {M.}~\bibnamefont
  {Aguena}}\ and\ \bibinfo {author} {\bibfnamefont {M.}~\bibnamefont {Lima}},\
  }\href {\doibase 10.1103/PhysRevD.98.123529} {\bibfield  {journal} {\bibinfo
  {journal} {Phys. Rev. D}\ }\textbf {\bibinfo {volume} {98}},\ \bibinfo
  {pages} {123529} (\bibinfo {year} {2018})},\ \Eprint
  {http://arxiv.org/abs/1611.05468} {arXiv:1611.05468 [astro-ph.CO]}
  \BibitemShut {NoStop}%
\bibitem [{\citenamefont {{Wen}}\ \emph {et~al.}(2020)\citenamefont {{Wen}},
  \citenamefont {{Kemball}},\ and\ \citenamefont {{Saslaw}}}]{cosd91}%
  \BibitemOpen
  \bibfield  {author} {\bibinfo {author} {\bibfnamefont {D.}~\bibnamefont
  {{Wen}}}, \bibinfo {author} {\bibfnamefont {A.~J.}\ \bibnamefont
  {{Kemball}}}, \ and\ \bibinfo {author} {\bibfnamefont {W.~C.}\ \bibnamefont
  {{Saslaw}}},\ }\href {\doibase 10.3847/1538-4357/ab6d6f} {\bibfield
  {journal} {\bibinfo  {journal} {Astrophys. J.}\ }\textbf {\bibinfo {volume}
  {890}},\ \bibinfo {eid} {160} (\bibinfo {year} {2020})},\ \Eprint
  {http://arxiv.org/abs/2001.06119} {arXiv:2001.06119 [astro-ph.CO]}
  \BibitemShut {NoStop}%
\bibitem [{\citenamefont {Hameeda}\ \emph {et~al.}(2016)\citenamefont
  {Hameeda}, \citenamefont {Upadhyay}, \citenamefont {Faizal},\ and\
  \citenamefont {Ali}}]{1b}%
  \BibitemOpen
  \bibfield  {author} {\bibinfo {author} {\bibfnamefont {M.}~\bibnamefont
  {Hameeda}}, \bibinfo {author} {\bibfnamefont {S.}~\bibnamefont {Upadhyay}},
  \bibinfo {author} {\bibfnamefont {M.}~\bibnamefont {Faizal}}, \ and\ \bibinfo
  {author} {\bibfnamefont {A.~F.}\ \bibnamefont {Ali}},\ }\href {\doibase
  10.1093/mnras/stw2202} {\bibfield  {journal} {\bibinfo  {journal} {Mon. Not.
  Roy. Astron. Soc.}\ }\textbf {\bibinfo {volume} {463}},\ \bibinfo {pages}
  {3699} (\bibinfo {year} {2016})},\ \Eprint {http://arxiv.org/abs/1610.07990}
  {arXiv:1610.07990 [gr-qc]} \BibitemShut {NoStop}%
\bibitem [{\citenamefont {Pourhassan}\ \emph {et~al.}(2017)\citenamefont
  {Pourhassan}, \citenamefont {Upadhyay}, \citenamefont {Hameeda},\ and\
  \citenamefont {Faizal}}]{1}%
  \BibitemOpen
  \bibfield  {author} {\bibinfo {author} {\bibfnamefont {B.}~\bibnamefont
  {Pourhassan}}, \bibinfo {author} {\bibfnamefont {S.}~\bibnamefont
  {Upadhyay}}, \bibinfo {author} {\bibfnamefont {M.}~\bibnamefont {Hameeda}}, \
  and\ \bibinfo {author} {\bibfnamefont {M.}~\bibnamefont {Faizal}},\ }\href
  {\doibase 10.1093/mnras/stx697} {\bibfield  {journal} {\bibinfo  {journal}
  {Mon. Not. Roy. Astron. Soc.}\ }\textbf {\bibinfo {volume} {468}},\ \bibinfo
  {pages} {3166} (\bibinfo {year} {2017})},\ \Eprint
  {http://arxiv.org/abs/1704.06085} {arXiv:1704.06085 [gr-qc]} \BibitemShut
  {NoStop}%
\bibitem [{\citenamefont {Deser}\ and\ \citenamefont {Woodard}(2007)}]{50acd}%
  \BibitemOpen
  \bibfield  {author} {\bibinfo {author} {\bibfnamefont {S.}~\bibnamefont
  {Deser}}\ and\ \bibinfo {author} {\bibfnamefont {R.~P.}\ \bibnamefont
  {Woodard}},\ }\href {\doibase 10.1103/PhysRevLett.99.111301} {\bibfield
  {journal} {\bibinfo  {journal} {Phys. Rev. Lett.}\ }\textbf {\bibinfo
  {volume} {99}},\ \bibinfo {pages} {111301} (\bibinfo {year} {2007})},\
  \Eprint {http://arxiv.org/abs/0706.2151} {arXiv:0706.2151 [astro-ph]}
  \BibitemShut {NoStop}%
\bibitem [{\citenamefont {{Ding}}\ and\ \citenamefont {{Deng}}(2019)}]{50ab}%
  \BibitemOpen
  \bibfield  {author} {\bibinfo {author} {\bibfnamefont {J.-C.}\ \bibnamefont
  {{Ding}}}\ and\ \bibinfo {author} {\bibfnamefont {J.-B.}\ \bibnamefont
  {{Deng}}},\ }\href {\doibase 10.1088/1475-7516/2019/12/054} {\bibfield
  {journal} {\bibinfo  {journal} {JCAP}\ }\textbf {\bibinfo {volume} {2019}},\
  \bibinfo {eid} {054} (\bibinfo {year} {2019})},\ \Eprint
  {http://arxiv.org/abs/1908.11223} {arXiv:1908.11223 [astro-ph.CO]}
  \BibitemShut {NoStop}%
\bibitem [{\citenamefont {Bahamonde}\ \emph
  {et~al.}(2017{\natexlab{a}})\citenamefont {Bahamonde}, \citenamefont
  {Capozziello},\ and\ \citenamefont {Dialektopoulos}}]{Bahamonde}%
  \BibitemOpen
  \bibfield  {author} {\bibinfo {author} {\bibfnamefont {S.}~\bibnamefont
  {Bahamonde}}, \bibinfo {author} {\bibfnamefont {S.}~\bibnamefont
  {Capozziello}}, \ and\ \bibinfo {author} {\bibfnamefont {K.~F.}\ \bibnamefont
  {Dialektopoulos}},\ }\href {\doibase 10.1140/epjc/s10052-017-5283-x}
  {\bibfield  {journal} {\bibinfo  {journal} {Eur. Phys. J. C}\ }\textbf
  {\bibinfo {volume} {77}},\ \bibinfo {pages} {722} (\bibinfo {year}
  {2017}{\natexlab{a}})},\ \Eprint {http://arxiv.org/abs/1708.06310}
  {arXiv:1708.06310 [gr-qc]} \BibitemShut {NoStop}%
\bibitem [{\citenamefont {Bahamonde}\ \emph
  {et~al.}(2017{\natexlab{b}})\citenamefont {Bahamonde}, \citenamefont
  {Capozziello}, \citenamefont {Faizal},\ and\ \citenamefont {Nunes}}]{Nunes}%
  \BibitemOpen
  \bibfield  {author} {\bibinfo {author} {\bibfnamefont {S.}~\bibnamefont
  {Bahamonde}}, \bibinfo {author} {\bibfnamefont {S.}~\bibnamefont
  {Capozziello}}, \bibinfo {author} {\bibfnamefont {M.}~\bibnamefont {Faizal}},
  \ and\ \bibinfo {author} {\bibfnamefont {R.~C.}\ \bibnamefont {Nunes}},\
  }\href {\doibase 10.1140/epjc/s10052-017-5210-1} {\bibfield  {journal}
  {\bibinfo  {journal} {Eur. Phys. J. C}\ }\textbf {\bibinfo {volume} {77}},\
  \bibinfo {pages} {628} (\bibinfo {year} {2017}{\natexlab{b}})},\ \Eprint
  {http://arxiv.org/abs/1709.02692} {arXiv:1709.02692 [gr-qc]} \BibitemShut
  {NoStop}%
\bibitem [{\citenamefont {Bajardi}\ \emph {et~al.}(2020)\citenamefont
  {Bajardi}, \citenamefont {Capozziello},\ and\ \citenamefont
  {Vernieri}}]{Bajardi}%
  \BibitemOpen
  \bibfield  {author} {\bibinfo {author} {\bibfnamefont {F.}~\bibnamefont
  {Bajardi}}, \bibinfo {author} {\bibfnamefont {S.}~\bibnamefont
  {Capozziello}}, \ and\ \bibinfo {author} {\bibfnamefont {D.}~\bibnamefont
  {Vernieri}},\ }\href {\doibase 10.1140/epjp/s13360-020-00944-1} {\bibfield
  {journal} {\bibinfo  {journal} {Eur. Phys. J. Plus}\ }\textbf {\bibinfo
  {volume} {135}},\ \bibinfo {pages} {942} (\bibinfo {year} {2020})},\ \Eprint
  {http://arxiv.org/abs/2011.01317} {arXiv:2011.01317 [gr-qc]} \BibitemShut
  {NoStop}%
\bibitem [{\citenamefont {Modesto}\ and\ \citenamefont
  {Rachwa\l{}}(2017)}]{Modesto}%
  \BibitemOpen
  \bibfield  {author} {\bibinfo {author} {\bibfnamefont {L.}~\bibnamefont
  {Modesto}}\ and\ \bibinfo {author} {\bibfnamefont {L.}~\bibnamefont
  {Rachwa\l{}}},\ }\href {\doibase 10.1142/S0218271817300208} {\bibfield
  {journal} {\bibinfo  {journal} {Int. J. Mod. Phys. D}\ }\textbf {\bibinfo
  {volume} {26}},\ \bibinfo {pages} {1730020} (\bibinfo {year}
  {2017})}\BibitemShut {NoStop}%
\bibitem [{\citenamefont {Capozziello}\ \emph {et~al.}(2020)\citenamefont
  {Capozziello}, \citenamefont {Capriolo},\ and\ \citenamefont
  {Nojiri}}]{Capriolo}%
  \BibitemOpen
  \bibfield  {author} {\bibinfo {author} {\bibfnamefont {S.}~\bibnamefont
  {Capozziello}}, \bibinfo {author} {\bibfnamefont {M.}~\bibnamefont
  {Capriolo}}, \ and\ \bibinfo {author} {\bibfnamefont {S.}~\bibnamefont
  {Nojiri}},\ }\href {\doibase 10.1016/j.physletb.2020.135821} {\bibfield
  {journal} {\bibinfo  {journal} {Phys. Lett. B}\ }\textbf {\bibinfo {volume}
  {810}},\ \bibinfo {pages} {135821} (\bibinfo {year} {2020})},\ \Eprint
  {http://arxiv.org/abs/2009.12777} {arXiv:2009.12777 [gr-qc]} \BibitemShut
  {NoStop}%
\bibitem [{\citenamefont {Belgacem}\ \emph {et~al.}(2019)\citenamefont
  {Belgacem}, \citenamefont {Dirian}, \citenamefont {Finke}, \citenamefont
  {Foffa},\ and\ \citenamefont {Maggiore}}]{50ac}%
  \BibitemOpen
  \bibfield  {author} {\bibinfo {author} {\bibfnamefont {E.}~\bibnamefont
  {Belgacem}}, \bibinfo {author} {\bibfnamefont {Y.}~\bibnamefont {Dirian}},
  \bibinfo {author} {\bibfnamefont {A.}~\bibnamefont {Finke}}, \bibinfo
  {author} {\bibfnamefont {S.}~\bibnamefont {Foffa}}, \ and\ \bibinfo {author}
  {\bibfnamefont {M.}~\bibnamefont {Maggiore}},\ }\href {\doibase
  10.1088/1475-7516/2019/11/022} {\bibfield  {journal} {\bibinfo  {journal}
  {JCAP}\ }\textbf {\bibinfo {volume} {11}},\ \bibinfo {pages} {022} (\bibinfo
  {year} {2019})},\ \Eprint {http://arxiv.org/abs/1907.02047} {arXiv:1907.02047
  [astro-ph.CO]} \BibitemShut {NoStop}%
\bibitem [{\citenamefont {Koshelev}(2007)}]{sftn1}%
  \BibitemOpen
  \bibfield  {author} {\bibinfo {author} {\bibfnamefont {A.~S.}\ \bibnamefont
  {Koshelev}},\ }\href {\doibase 10.1088/1126-6708/2007/04/029} {\bibfield
  {journal} {\bibinfo  {journal} {JHEP}\ }\textbf {\bibinfo {volume} {04}},\
  \bibinfo {pages} {029} (\bibinfo {year} {2007})},\ \Eprint
  {http://arxiv.org/abs/hep-th/0701103} {arXiv:hep-th/0701103} \BibitemShut
  {NoStop}%
\bibitem [{\citenamefont {Koshelev}\ and\ \citenamefont
  {Vernov}(2012)}]{sftn2}%
  \BibitemOpen
  \bibfield  {author} {\bibinfo {author} {\bibfnamefont {A.~S.}\ \bibnamefont
  {Koshelev}}\ and\ \bibinfo {author} {\bibfnamefont {S.~Y.}\ \bibnamefont
  {Vernov}},\ }\href {\doibase 10.1140/epjc/s10052-012-2198-4} {\bibfield
  {journal} {\bibinfo  {journal} {Eur. Phys. J. C}\ }\textbf {\bibinfo {volume}
  {72}},\ \bibinfo {pages} {2198} (\bibinfo {year} {2012})},\ \Eprint
  {http://arxiv.org/abs/0903.5176} {arXiv:0903.5176 [hep-th]} \BibitemShut
  {NoStop}%
\bibitem [{\citenamefont {Gielen}\ \emph {et~al.}(2013)\citenamefont {Gielen},
  \citenamefont {Oriti},\ and\ \citenamefont {Sindoni}}]{lqgn1}%
  \BibitemOpen
  \bibfield  {author} {\bibinfo {author} {\bibfnamefont {S.}~\bibnamefont
  {Gielen}}, \bibinfo {author} {\bibfnamefont {D.}~\bibnamefont {Oriti}}, \
  and\ \bibinfo {author} {\bibfnamefont {L.}~\bibnamefont {Sindoni}},\ }\href
  {\doibase 10.1103/PhysRevLett.111.031301} {\bibfield  {journal} {\bibinfo
  {journal} {Phys. Rev. Lett.}\ }\textbf {\bibinfo {volume} {111}},\ \bibinfo
  {pages} {031301} (\bibinfo {year} {2013})},\ \Eprint
  {http://arxiv.org/abs/1303.3576} {arXiv:1303.3576 [gr-qc]} \BibitemShut
  {NoStop}%
\bibitem [{\citenamefont {Gielen}(2019)}]{lqgn2}%
  \BibitemOpen
  \bibfield  {author} {\bibinfo {author} {\bibfnamefont {S.}~\bibnamefont
  {Gielen}},\ }\href {\doibase 10.1088/1475-7516/2019/02/013} {\bibfield
  {journal} {\bibinfo  {journal} {JCAP}\ }\textbf {\bibinfo {volume} {02}},\
  \bibinfo {pages} {013} (\bibinfo {year} {2019})},\ \Eprint
  {http://arxiv.org/abs/1811.10639} {arXiv:1811.10639 [gr-qc]} \BibitemShut
  {NoStop}%
\bibitem [{\citenamefont {Yunes}\ and\ \citenamefont {Siemens}(2013)}]{50}%
  \BibitemOpen
  \bibfield  {author} {\bibinfo {author} {\bibfnamefont {N.}~\bibnamefont
  {Yunes}}\ and\ \bibinfo {author} {\bibfnamefont {X.}~\bibnamefont
  {Siemens}},\ }\href {\doibase 10.12942/lrr-2013-9} {\bibfield  {journal}
  {\bibinfo  {journal} {Living Rev. Rel.}\ }\textbf {\bibinfo {volume} {16}},\
  \bibinfo {pages} {9} (\bibinfo {year} {2013})},\ \Eprint
  {http://arxiv.org/abs/1304.3473} {arXiv:1304.3473 [gr-qc]} \BibitemShut
  {NoStop}%
\bibitem [{\citenamefont {Chicone}\ and\ \citenamefont
  {Mashhoon}(2016{\natexlab{a}})}]{50a}%
  \BibitemOpen
  \bibfield  {author} {\bibinfo {author} {\bibfnamefont {C.}~\bibnamefont
  {Chicone}}\ and\ \bibinfo {author} {\bibfnamefont {B.}~\bibnamefont
  {Mashhoon}},\ }\href {\doibase 10.1088/0264-9381/33/7/075005} {\bibfield
  {journal} {\bibinfo  {journal} {Class. Quant. Grav.}\ }\textbf {\bibinfo
  {volume} {33}},\ \bibinfo {pages} {075005} (\bibinfo {year}
  {2016}{\natexlab{a}})},\ \Eprint {http://arxiv.org/abs/1508.01508}
  {arXiv:1508.01508 [gr-qc]} \BibitemShut {NoStop}%
\bibitem [{\citenamefont {{Tian}}\ and\ \citenamefont {{Zhu}}(2019)}]{50b}%
  \BibitemOpen
  \bibfield  {author} {\bibinfo {author} {\bibfnamefont {S.~X.}\ \bibnamefont
  {{Tian}}}\ and\ \bibinfo {author} {\bibfnamefont {Z.-H.}\ \bibnamefont
  {{Zhu}}},\ }\href {\doibase 10.1103/PhysRevD.100.124059} {\bibfield
  {journal} {\bibinfo  {journal} {Phys. Rev. D}\ }\textbf {\bibinfo {volume}
  {100}},\ \bibinfo {eid} {124059} (\bibinfo {year} {2019})}\BibitemShut
  {NoStop}%
\bibitem [{\citenamefont {Dialektopoulos}\ \emph {et~al.}(2019)\citenamefont
  {Dialektopoulos}, \citenamefont {Borka}, \citenamefont {Capozziello},
  \citenamefont {Borka~Jovanovi\'c},\ and\ \citenamefont
  {Jovanovi\'c}}]{Kostas}%
  \BibitemOpen
  \bibfield  {author} {\bibinfo {author} {\bibfnamefont {K.~F.}\ \bibnamefont
  {Dialektopoulos}}, \bibinfo {author} {\bibfnamefont {D.}~\bibnamefont
  {Borka}}, \bibinfo {author} {\bibfnamefont {S.}~\bibnamefont {Capozziello}},
  \bibinfo {author} {\bibfnamefont {V.}~\bibnamefont {Borka~Jovanovi\'c}}, \
  and\ \bibinfo {author} {\bibfnamefont {P.}~\bibnamefont {Jovanovi\'c}},\
  }\href {\doibase 10.1103/PhysRevD.99.044053} {\bibfield  {journal} {\bibinfo
  {journal} {Phys. Rev. D}\ }\textbf {\bibinfo {volume} {99}},\ \bibinfo
  {pages} {044053} (\bibinfo {year} {2019})},\ \Eprint
  {http://arxiv.org/abs/1812.09289} {arXiv:1812.09289 [astro-ph.GA]}
  \BibitemShut {NoStop}%
\bibitem [{\citenamefont {Lazar}(2020)}]{Lazar:2020gsx}%
  \BibitemOpen
  \bibfield  {author} {\bibinfo {author} {\bibfnamefont {M.}~\bibnamefont
  {Lazar}},\ }\href {\doibase 10.1103/PhysRevD.102.096002} {\bibfield
  {journal} {\bibinfo  {journal} {Phys. Rev. D}\ }\textbf {\bibinfo {volume}
  {102}},\ \bibinfo {pages} {096002} (\bibinfo {year} {2020})},\ \Eprint
  {http://arxiv.org/abs/2009.09846} {arXiv:2009.09846 [gr-qc]} \BibitemShut
  {NoStop}%
\bibitem [{\citenamefont {Amendola}\ \emph {et~al.}(2019)\citenamefont
  {Amendola}, \citenamefont {Dirian}, \citenamefont {Nersisyan},\ and\
  \citenamefont {Park}}]{2019JCAP}%
  \BibitemOpen
  \bibfield  {author} {\bibinfo {author} {\bibfnamefont {L.}~\bibnamefont
  {Amendola}}, \bibinfo {author} {\bibfnamefont {Y.}~\bibnamefont {Dirian}},
  \bibinfo {author} {\bibfnamefont {H.}~\bibnamefont {Nersisyan}}, \ and\
  \bibinfo {author} {\bibfnamefont {S.}~\bibnamefont {Park}},\ }\href {\doibase
  10.1088/1475-7516/2019/03/045} {\bibfield  {journal} {\bibinfo  {journal}
  {JCAP}\ }\textbf {\bibinfo {volume} {03}},\ \bibinfo {pages} {045} (\bibinfo
  {year} {2019})},\ \Eprint {http://arxiv.org/abs/1901.07832} {arXiv:1901.07832
  [astro-ph.CO]} \BibitemShut {NoStop}%
\bibitem [{\citenamefont {Hehl}\ and\ \citenamefont
  {Mashhoon}(2009{\natexlab{a}})}]{40a}%
  \BibitemOpen
  \bibfield  {author} {\bibinfo {author} {\bibfnamefont {F.~W.}\ \bibnamefont
  {Hehl}}\ and\ \bibinfo {author} {\bibfnamefont {B.}~\bibnamefont
  {Mashhoon}},\ }\href {\doibase 10.1103/PhysRevD.79.064028} {\bibfield
  {journal} {\bibinfo  {journal} {Phys. Rev. D}\ }\textbf {\bibinfo {volume}
  {79}},\ \bibinfo {pages} {064028} (\bibinfo {year} {2009}{\natexlab{a}})},\
  \Eprint {http://arxiv.org/abs/0902.0560} {arXiv:0902.0560 [gr-qc]}
  \BibitemShut {NoStop}%
\bibitem [{\citenamefont {Hehl}\ and\ \citenamefont
  {Mashhoon}(2009{\natexlab{b}})}]{40b}%
  \BibitemOpen
  \bibfield  {author} {\bibinfo {author} {\bibfnamefont {F.~W.}\ \bibnamefont
  {Hehl}}\ and\ \bibinfo {author} {\bibfnamefont {B.}~\bibnamefont
  {Mashhoon}},\ }\href {\doibase 10.1016/j.physletb.2009.02.033} {\bibfield
  {journal} {\bibinfo  {journal} {Phys. Lett. B}\ }\textbf {\bibinfo {volume}
  {673}},\ \bibinfo {pages} {279} (\bibinfo {year} {2009}{\natexlab{b}})},\
  \Eprint {http://arxiv.org/abs/0812.1059} {arXiv:0812.1059 [gr-qc]}
  \BibitemShut {NoStop}%
\bibitem [{\citenamefont {Blome}\ \emph {et~al.}(2010)\citenamefont {Blome},
  \citenamefont {Chicone}, \citenamefont {Hehl},\ and\ \citenamefont
  {Mashhoon}}]{40c}%
  \BibitemOpen
  \bibfield  {author} {\bibinfo {author} {\bibfnamefont {H.-J.}\ \bibnamefont
  {Blome}}, \bibinfo {author} {\bibfnamefont {C.}~\bibnamefont {Chicone}},
  \bibinfo {author} {\bibfnamefont {F.~W.}\ \bibnamefont {Hehl}}, \ and\
  \bibinfo {author} {\bibfnamefont {B.}~\bibnamefont {Mashhoon}},\ }\href
  {\doibase 10.1103/PhysRevD.81.065020} {\bibfield  {journal} {\bibinfo
  {journal} {Phys. Rev. D}\ }\textbf {\bibinfo {volume} {81}},\ \bibinfo
  {pages} {065020} (\bibinfo {year} {2010})},\ \Eprint
  {http://arxiv.org/abs/1002.1425} {arXiv:1002.1425 [gr-qc]} \BibitemShut
  {NoStop}%
\bibitem [{\citenamefont {Chicone}\ and\ \citenamefont
  {Mashhoon}(2016{\natexlab{b}})}]{40d}%
  \BibitemOpen
  \bibfield  {author} {\bibinfo {author} {\bibfnamefont {C.}~\bibnamefont
  {Chicone}}\ and\ \bibinfo {author} {\bibfnamefont {B.}~\bibnamefont
  {Mashhoon}},\ }\href {\doibase 10.1063/1.4958902} {\bibfield  {journal}
  {\bibinfo  {journal} {J. Math. Phys.}\ }\textbf {\bibinfo {volume} {57}},\
  \bibinfo {pages} {072501} (\bibinfo {year} {2016}{\natexlab{b}})},\ \Eprint
  {http://arxiv.org/abs/1510.07316} {arXiv:1510.07316 [gr-qc]} \BibitemShut
  {NoStop}%
\bibitem [{\citenamefont {Shapiro}\ \emph {et~al.}(2005)\citenamefont
  {Shapiro}, \citenamefont {Sola},\ and\ \citenamefont
  {Stefancic}}]{Shapiro:2004ch}%
  \BibitemOpen
  \bibfield  {author} {\bibinfo {author} {\bibfnamefont {I.~L.}\ \bibnamefont
  {Shapiro}}, \bibinfo {author} {\bibfnamefont {J.}~\bibnamefont {Sola}}, \
  and\ \bibinfo {author} {\bibfnamefont {H.}~\bibnamefont {Stefancic}},\ }\href
  {\doibase 10.1088/1475-7516/2005/01/012} {\bibfield  {journal} {\bibinfo
  {journal} {JCAP}\ }\textbf {\bibinfo {volume} {01}},\ \bibinfo {pages} {012}
  (\bibinfo {year} {2005})},\ \Eprint {http://arxiv.org/abs/hep-ph/0410095}
  {arXiv:hep-ph/0410095} \BibitemShut {NoStop}%
\bibitem [{\citenamefont {Famaey}\ and\ \citenamefont
  {McGaugh}(2012)}]{Famaey:2011kh}%
  \BibitemOpen
  \bibfield  {author} {\bibinfo {author} {\bibfnamefont {B.}~\bibnamefont
  {Famaey}}\ and\ \bibinfo {author} {\bibfnamefont {S.}~\bibnamefont
  {McGaugh}},\ }\href {\doibase 10.12942/lrr-2012-10} {\bibfield  {journal}
  {\bibinfo  {journal} {Living Rev. Rel.}\ }\textbf {\bibinfo {volume} {15}},\
  \bibinfo {pages} {10} (\bibinfo {year} {2012})},\ \Eprint
  {http://arxiv.org/abs/1112.3960} {arXiv:1112.3960 [astro-ph.CO]} \BibitemShut
  {NoStop}%
\bibitem [{\citenamefont {Park}\ and\ \citenamefont
  {Tamaryan}(2003)}]{Park:2002rb}%
  \BibitemOpen
  \bibfield  {author} {\bibinfo {author} {\bibfnamefont {D.~K.}\ \bibnamefont
  {Park}}\ and\ \bibinfo {author} {\bibfnamefont {S.}~\bibnamefont
  {Tamaryan}},\ }\href {\doibase 10.1016/S0370-2693(02)03269-0} {\bibfield
  {journal} {\bibinfo  {journal} {Phys. Lett. B}\ }\textbf {\bibinfo {volume}
  {554}},\ \bibinfo {pages} {92} (\bibinfo {year} {2003})},\ \Eprint
  {http://arxiv.org/abs/hep-th/0212023} {arXiv:hep-th/0212023} \BibitemShut
  {NoStop}%
\bibitem [{\citenamefont {Soleng}(1995)}]{Soleng:1993yr}%
  \BibitemOpen
  \bibfield  {author} {\bibinfo {author} {\bibfnamefont {H.~H.}\ \bibnamefont
  {Soleng}},\ }\href {\doibase 10.1007/BF02107935} {\bibfield  {journal}
  {\bibinfo  {journal} {Gen. Rel. Grav.}\ }\textbf {\bibinfo {volume} {27}},\
  \bibinfo {pages} {367} (\bibinfo {year} {1995})},\ \Eprint
  {http://arxiv.org/abs/gr-qc/9412053} {arXiv:gr-qc/9412053} \BibitemShut
  {NoStop}%
\bibitem [{\citenamefont {Fabris}\ and\ \citenamefont
  {Campos}(2009)}]{Fabris:2007df}%
  \BibitemOpen
  \bibfield  {author} {\bibinfo {author} {\bibfnamefont {J.~C.}\ \bibnamefont
  {Fabris}}\ and\ \bibinfo {author} {\bibfnamefont {J.~P.}\ \bibnamefont
  {Campos}},\ }\href {\doibase 10.1007/s10714-008-0654-0} {\bibfield  {journal}
  {\bibinfo  {journal} {Gen. Rel. Grav.}\ }\textbf {\bibinfo {volume} {41}},\
  \bibinfo {pages} {93} (\bibinfo {year} {2009})},\ \Eprint
  {http://arxiv.org/abs/0710.3683} {arXiv:0710.3683 [astro-ph]} \BibitemShut
  {NoStop}%
\bibitem [{\citenamefont {Calcagni}\ and\ \citenamefont
  {Modesto}(2014)}]{Calcagni:2013eua}%
  \BibitemOpen
  \bibfield  {author} {\bibinfo {author} {\bibfnamefont {G.}~\bibnamefont
  {Calcagni}}\ and\ \bibinfo {author} {\bibfnamefont {L.}~\bibnamefont
  {Modesto}},\ }\href {\doibase 10.1088/1751-8113/47/35/355402} {\bibfield
  {journal} {\bibinfo  {journal} {J. Phys. A}\ }\textbf {\bibinfo {volume}
  {47}},\ \bibinfo {pages} {355402} (\bibinfo {year} {2014})},\ \Eprint
  {http://arxiv.org/abs/1310.4957} {arXiv:1310.4957 [hep-th]} \BibitemShut
  {NoStop}%
\bibitem [{\citenamefont {Biswas}\ \emph {et~al.}(2012)\citenamefont {Biswas},
  \citenamefont {Kapusta},\ and\ \citenamefont {Reddy}}]{Biswas:2012ka}%
  \BibitemOpen
  \bibfield  {author} {\bibinfo {author} {\bibfnamefont {T.}~\bibnamefont
  {Biswas}}, \bibinfo {author} {\bibfnamefont {J.}~\bibnamefont {Kapusta}}, \
  and\ \bibinfo {author} {\bibfnamefont {A.}~\bibnamefont {Reddy}},\ }\href
  {\doibase 10.1007/JHEP12(2012)008} {\bibfield  {journal} {\bibinfo  {journal}
  {JHEP}\ }\textbf {\bibinfo {volume} {12}},\ \bibinfo {pages} {008} (\bibinfo
  {year} {2012})},\ \Eprint {http://arxiv.org/abs/1201.1580} {arXiv:1201.1580
  [hep-th]} \BibitemShut {NoStop}%
\bibitem [{\citenamefont {Soussa}\ and\ \citenamefont
  {Woodard}(2003)}]{Soussa:2003vv}%
  \BibitemOpen
  \bibfield  {author} {\bibinfo {author} {\bibfnamefont {M.~E.}\ \bibnamefont
  {Soussa}}\ and\ \bibinfo {author} {\bibfnamefont {R.~P.}\ \bibnamefont
  {Woodard}},\ }\href {\doibase 10.1088/0264-9381/20/13/321} {\bibfield
  {journal} {\bibinfo  {journal} {Class. Quant. Grav.}\ }\textbf {\bibinfo
  {volume} {20}},\ \bibinfo {pages} {2737} (\bibinfo {year} {2003})},\ \Eprint
  {http://arxiv.org/abs/astro-ph/0302030} {arXiv:astro-ph/0302030} \BibitemShut
  {NoStop}%
\bibitem [{\citenamefont {Deffayet}\ \emph {et~al.}(2011)\citenamefont
  {Deffayet}, \citenamefont {Esposito-Farese},\ and\ \citenamefont
  {Woodard}}]{Deffayet:2011sk}%
  \BibitemOpen
  \bibfield  {author} {\bibinfo {author} {\bibfnamefont {C.}~\bibnamefont
  {Deffayet}}, \bibinfo {author} {\bibfnamefont {G.}~\bibnamefont
  {Esposito-Farese}}, \ and\ \bibinfo {author} {\bibfnamefont {R.~P.}\
  \bibnamefont {Woodard}},\ }\href {\doibase 10.1103/PhysRevD.84.124054}
  {\bibfield  {journal} {\bibinfo  {journal} {Phys. Rev. D}\ }\textbf {\bibinfo
  {volume} {84}},\ \bibinfo {pages} {124054} (\bibinfo {year} {2011})},\
  \Eprint {http://arxiv.org/abs/1106.4984} {arXiv:1106.4984 [gr-qc]}
  \BibitemShut {NoStop}%
\bibitem [{\citenamefont {Deffayet}\ \emph {et~al.}(2014)\citenamefont
  {Deffayet}, \citenamefont {Esposito-Farese},\ and\ \citenamefont
  {Woodard}}]{Deffayet:2014lba}%
  \BibitemOpen
  \bibfield  {author} {\bibinfo {author} {\bibfnamefont {C.}~\bibnamefont
  {Deffayet}}, \bibinfo {author} {\bibfnamefont {G.}~\bibnamefont
  {Esposito-Farese}}, \ and\ \bibinfo {author} {\bibfnamefont {R.~P.}\
  \bibnamefont {Woodard}},\ }\href {\doibase 10.1103/PhysRevD.90.089901}
  {\bibfield  {journal} {\bibinfo  {journal} {Phys. Rev. D}\ }\textbf {\bibinfo
  {volume} {90}},\ \bibinfo {pages} {064038} (\bibinfo {year} {2014})},\
  \bibinfo {note} {[Addendum: Phys.Rev.D 90, 089901 (2014)]},\ \Eprint
  {http://arxiv.org/abs/1405.0393} {arXiv:1405.0393 [astro-ph.CO]} \BibitemShut
  {NoStop}%
\bibitem [{\citenamefont {Tan}\ and\ \citenamefont
  {Woodard}(2018)}]{Tan:2018bfp}%
  \BibitemOpen
  \bibfield  {author} {\bibinfo {author} {\bibfnamefont {L.}~\bibnamefont
  {Tan}}\ and\ \bibinfo {author} {\bibfnamefont {R.~P.}\ \bibnamefont
  {Woodard}},\ }\href {\doibase 10.1088/1475-7516/2018/05/037} {\bibfield
  {journal} {\bibinfo  {journal} {JCAP}\ }\textbf {\bibinfo {volume} {05}},\
  \bibinfo {pages} {037} (\bibinfo {year} {2018})},\ \Eprint
  {http://arxiv.org/abs/1804.01669} {arXiv:1804.01669 [gr-qc]} \BibitemShut
  {NoStop}%
\bibitem [{\citenamefont {Kim}\ \emph {et~al.}(2016)\citenamefont {Kim},
  \citenamefont {Rahat}, \citenamefont {Sayeb}, \citenamefont {Tan},
  \citenamefont {Woodard},\ and\ \citenamefont {Xu}}]{Kim:2016nnd}%
  \BibitemOpen
  \bibfield  {author} {\bibinfo {author} {\bibfnamefont {M.}~\bibnamefont
  {Kim}}, \bibinfo {author} {\bibfnamefont {M.~H.}\ \bibnamefont {Rahat}},
  \bibinfo {author} {\bibfnamefont {M.}~\bibnamefont {Sayeb}}, \bibinfo
  {author} {\bibfnamefont {L.}~\bibnamefont {Tan}}, \bibinfo {author}
  {\bibfnamefont {R.~P.}\ \bibnamefont {Woodard}}, \ and\ \bibinfo {author}
  {\bibfnamefont {B.}~\bibnamefont {Xu}},\ }\href {\doibase
  10.1103/PhysRevD.94.104009} {\bibfield  {journal} {\bibinfo  {journal} {Phys.
  Rev. D}\ }\textbf {\bibinfo {volume} {94}},\ \bibinfo {pages} {104009}
  (\bibinfo {year} {2016})},\ \Eprint {http://arxiv.org/abs/1608.07858}
  {arXiv:1608.07858 [gr-qc]} \BibitemShut {NoStop}%
\bibitem [{\citenamefont {Bagchi}\ and\ \citenamefont
  {Fring}(2019)}]{Bagchi:2017jfl}%
  \BibitemOpen
  \bibfield  {author} {\bibinfo {author} {\bibfnamefont {B.}~\bibnamefont
  {Bagchi}}\ and\ \bibinfo {author} {\bibfnamefont {A.}~\bibnamefont {Fring}},\
  }\href {\doibase 10.1142/S0217979219500188} {\bibfield  {journal} {\bibinfo
  {journal} {Int. J. Mod. Phys. B}\ }\textbf {\bibinfo {volume} {33}},\
  \bibinfo {pages} {1950018} (\bibinfo {year} {2019})},\ \Eprint
  {http://arxiv.org/abs/1709.04339} {arXiv:1709.04339 [physics.gen-ph]}
  \BibitemShut {NoStop}%
\bibitem [{\citenamefont {{Wen}}\ \emph {et~al.}(2012)\citenamefont {{Wen}},
  \citenamefont {{Han}},\ and\ \citenamefont {{Liu}}}]{WenHan2012}%
  \BibitemOpen
  \bibfield  {author} {\bibinfo {author} {\bibfnamefont {Z.~L.}\ \bibnamefont
  {{Wen}}}, \bibinfo {author} {\bibfnamefont {J.~L.}\ \bibnamefont {{Han}}}, \
  and\ \bibinfo {author} {\bibfnamefont {F.~S.}\ \bibnamefont {{Liu}}},\ }\href
  {\doibase 10.1088/0067-0049/199/2/34} {\bibfield  {journal} {\bibinfo
  {journal} {Astrophysical J. Supplement}\ }\textbf {\bibinfo {volume} {199}},\
  \bibinfo {eid} {34} (\bibinfo {year} {2012})},\ \Eprint
  {http://arxiv.org/abs/1202.6424} {arXiv:1202.6424 [astro-ph.CO]} \BibitemShut
  {NoStop}%
\bibitem [{\citenamefont {{Wen}}\ and\ \citenamefont
  {{Han}}(2015)}]{WenHan2015}%
  \BibitemOpen
  \bibfield  {author} {\bibinfo {author} {\bibfnamefont {Z.~L.}\ \bibnamefont
  {{Wen}}}\ and\ \bibinfo {author} {\bibfnamefont {J.~L.}\ \bibnamefont
  {{Han}}},\ }\href {\doibase 10.1088/0004-637X/807/2/178} {\bibfield
  {journal} {\bibinfo  {journal} {Astrophys. J.}\ }\textbf {\bibinfo {volume}
  {807}},\ \bibinfo {eid} {178} (\bibinfo {year} {2015})},\ \Eprint
  {http://arxiv.org/abs/1506.04503} {arXiv:1506.04503 [astro-ph.GA]}
  \BibitemShut {NoStop}%
\bibitem [{\citenamefont {Yang}\ and\ \citenamefont {Saslaw}(2011)}]{Yang11}%
  \BibitemOpen
  \bibfield  {author} {\bibinfo {author} {\bibfnamefont {A.}~\bibnamefont
  {Yang}}\ and\ \bibinfo {author} {\bibfnamefont {W.~C.}\ \bibnamefont
  {Saslaw}},\ }\href {\doibase 10.1088/0004-637X/729/2/123} {\bibfield
  {journal} {\bibinfo  {journal} {Astrophys. J.}\ }\textbf {\bibinfo {volume}
  {729}},\ \bibinfo {pages} {123} (\bibinfo {year} {2011})},\ \Eprint
  {http://arxiv.org/abs/1009.0013} {arXiv:1009.0013 [astro-ph.CO]} \BibitemShut
  {NoStop}%
\bibitem [{\citenamefont {Yang}\ and\ \citenamefont {Saslaw}(2012)}]{Yang12}%
  \BibitemOpen
  \bibfield  {author} {\bibinfo {author} {\bibfnamefont {A.}~\bibnamefont
  {Yang}}\ and\ \bibinfo {author} {\bibfnamefont {W.~C.}\ \bibnamefont
  {Saslaw}},\ }\href {\doibase 10.1088/0004-637X/753/2/113} {\bibfield
  {journal} {\bibinfo  {journal} {Astrophys. J.}\ }\textbf {\bibinfo {volume}
  {753}},\ \bibinfo {pages} {113} (\bibinfo {year} {2012})},\ \Eprint
  {http://arxiv.org/abs/1203.3823} {arXiv:1203.3823 [astro-ph.CO]} \BibitemShut
  {NoStop}%
\bibitem [{\citenamefont {Capozziello}(2002)}]{Curvature}%
  \BibitemOpen
  \bibfield  {author} {\bibinfo {author} {\bibfnamefont {S.}~\bibnamefont
  {Capozziello}},\ }\href {\doibase 10.1142/S0218271802002025} {\bibfield
  {journal} {\bibinfo  {journal} {Int. J. Mod. Phys. D}\ }\textbf {\bibinfo
  {volume} {11}},\ \bibinfo {pages} {483} (\bibinfo {year} {2002})},\ \Eprint
  {http://arxiv.org/abs/gr-qc/0201033} {arXiv:gr-qc/0201033} \BibitemShut
  {NoStop}%
\bibitem [{\citenamefont {Capozziello}\ and\ \citenamefont
  {De~Laurentis}(2011)}]{Report}%
  \BibitemOpen
  \bibfield  {author} {\bibinfo {author} {\bibfnamefont {S.}~\bibnamefont
  {Capozziello}}\ and\ \bibinfo {author} {\bibfnamefont {M.}~\bibnamefont
  {De~Laurentis}},\ }\href {\doibase 10.1016/j.physrep.2011.09.003} {\bibfield
  {journal} {\bibinfo  {journal} {Phys. Rept.}\ }\textbf {\bibinfo {volume}
  {509}},\ \bibinfo {pages} {167} (\bibinfo {year} {2011})},\ \Eprint
  {http://arxiv.org/abs/1108.6266} {arXiv:1108.6266 [gr-qc]} \BibitemShut
  {NoStop}%
\bibitem [{\citenamefont {Sotiriou}\ and\ \citenamefont
  {Faraoni}(2010)}]{gravity}%
  \BibitemOpen
  \bibfield  {author} {\bibinfo {author} {\bibfnamefont {T.~P.}\ \bibnamefont
  {Sotiriou}}\ and\ \bibinfo {author} {\bibfnamefont {V.}~\bibnamefont
  {Faraoni}},\ }\href {\doibase 10.1103/RevModPhys.82.451} {\bibfield
  {journal} {\bibinfo  {journal} {Rev. Mod. Phys.}\ }\textbf {\bibinfo {volume}
  {82}},\ \bibinfo {pages} {451} (\bibinfo {year} {2010})},\ \Eprint
  {http://arxiv.org/abs/0805.1726} {arXiv:0805.1726 [gr-qc]} \BibitemShut
  {NoStop}%
\bibitem [{\citenamefont {Nojiri}\ \emph {et~al.}(2017)\citenamefont {Nojiri},
  \citenamefont {Odintsov},\ and\ \citenamefont {Oikonomou}}]{Oikonomou}%
  \BibitemOpen
  \bibfield  {author} {\bibinfo {author} {\bibfnamefont {S.}~\bibnamefont
  {Nojiri}}, \bibinfo {author} {\bibfnamefont {S.~D.}\ \bibnamefont
  {Odintsov}}, \ and\ \bibinfo {author} {\bibfnamefont {V.~K.}\ \bibnamefont
  {Oikonomou}},\ }\href {\doibase 10.1016/j.physrep.2017.06.001} {\bibfield
  {journal} {\bibinfo  {journal} {Phys. Rept.}\ }\textbf {\bibinfo {volume}
  {692}},\ \bibinfo {pages} {1} (\bibinfo {year} {2017})},\ \Eprint
  {http://arxiv.org/abs/1705.11098} {arXiv:1705.11098 [gr-qc]} \BibitemShut
  {NoStop}%
\bibitem [{\citenamefont {Capozziello}\ \emph {et~al.}(2009)\citenamefont
  {Capozziello}, \citenamefont {De~Filippis},\ and\ \citenamefont
  {Salzano}}]{Betty}%
  \BibitemOpen
  \bibfield  {author} {\bibinfo {author} {\bibfnamefont {S.}~\bibnamefont
  {Capozziello}}, \bibinfo {author} {\bibfnamefont {E.}~\bibnamefont
  {De~Filippis}}, \ and\ \bibinfo {author} {\bibfnamefont {V.}~\bibnamefont
  {Salzano}},\ }\href {\doibase 10.1111/j.1365-2966.2008.14382.x} {\bibfield
  {journal} {\bibinfo  {journal} {Mon. Not. Roy. Astron. Soc.}\ }\textbf
  {\bibinfo {volume} {394}},\ \bibinfo {pages} {947} (\bibinfo {year}
  {2009})},\ \Eprint {http://arxiv.org/abs/0809.1882} {arXiv:0809.1882
  [astro-ph]} \BibitemShut {NoStop}%
\bibitem [{\citenamefont {Capozziello}\ \emph {et~al.}(2018)\citenamefont
  {Capozziello}, \citenamefont {Faizal}, \citenamefont {Hameeda}, \citenamefont
  {Pourhassan}, \citenamefont {Salzano},\ and\ \citenamefont
  {Upadhyay}}]{Cap2}%
  \BibitemOpen
  \bibfield  {author} {\bibinfo {author} {\bibfnamefont {S.}~\bibnamefont
  {Capozziello}}, \bibinfo {author} {\bibfnamefont {M.}~\bibnamefont {Faizal}},
  \bibinfo {author} {\bibfnamefont {M.}~\bibnamefont {Hameeda}}, \bibinfo
  {author} {\bibfnamefont {B.}~\bibnamefont {Pourhassan}}, \bibinfo {author}
  {\bibfnamefont {V.}~\bibnamefont {Salzano}}, \ and\ \bibinfo {author}
  {\bibfnamefont {S.}~\bibnamefont {Upadhyay}},\ }\href {\doibase
  10.1093/mnras/stx2945} {\bibfield  {journal} {\bibinfo  {journal} {Mon. Not.
  Roy. Astron. Soc.}\ }\textbf {\bibinfo {volume} {474}},\ \bibinfo {pages}
  {2430} (\bibinfo {year} {2018})},\ \Eprint {http://arxiv.org/abs/1711.06630}
  {arXiv:1711.06630 [astro-ph.CO]} \BibitemShut {NoStop}%
\bibitem [{\citenamefont {De~Martino}\ \emph {et~al.}(2014)\citenamefont
  {De~Martino}, \citenamefont {De~Laurentis}, \citenamefont {Atrio-Barandela},\
  and\ \citenamefont {Capozziello}}]{2az}%
  \BibitemOpen
  \bibfield  {author} {\bibinfo {author} {\bibfnamefont {I.}~\bibnamefont
  {De~Martino}}, \bibinfo {author} {\bibfnamefont {M.}~\bibnamefont
  {De~Laurentis}}, \bibinfo {author} {\bibfnamefont {F.}~\bibnamefont
  {Atrio-Barandela}}, \ and\ \bibinfo {author} {\bibfnamefont {S.}~\bibnamefont
  {Capozziello}},\ }\href {\doibase 10.1093/mnras/stu903} {\bibfield  {journal}
  {\bibinfo  {journal} {Mon. Not. Roy. Astron. Soc.}\ }\textbf {\bibinfo
  {volume} {442}},\ \bibinfo {pages} {921} (\bibinfo {year} {2014})},\ \Eprint
  {http://arxiv.org/abs/1310.0693} {arXiv:1310.0693 [astro-ph.CO]} \BibitemShut
  {NoStop}%
\bibitem [{\citenamefont {Hameeda}\ \emph {et~al.}(2019)\citenamefont
  {Hameeda}, \citenamefont {Pourhassan}, \citenamefont {Faizal}, \citenamefont
  {Masroor}, \citenamefont {Ansari},\ and\ \citenamefont {Suresh}}]{4}%
  \BibitemOpen
  \bibfield  {author} {\bibinfo {author} {\bibfnamefont {M.}~\bibnamefont
  {Hameeda}}, \bibinfo {author} {\bibfnamefont {B.}~\bibnamefont {Pourhassan}},
  \bibinfo {author} {\bibfnamefont {M.}~\bibnamefont {Faizal}}, \bibinfo
  {author} {\bibfnamefont {C.~P.}\ \bibnamefont {Masroor}}, \bibinfo {author}
  {\bibfnamefont {R.~U.~H.}\ \bibnamefont {Ansari}}, \ and\ \bibinfo {author}
  {\bibfnamefont {P.~K.}\ \bibnamefont {Suresh}},\ }\href {\doibase
  10.1140/epjc/s10052-019-7281-7} {\bibfield  {journal} {\bibinfo  {journal}
  {Eur. Phys. J. C}\ }\textbf {\bibinfo {volume} {79}},\ \bibinfo {pages} {769}
  (\bibinfo {year} {2019})},\ \Eprint {http://arxiv.org/abs/1911.01739}
  {arXiv:1911.01739 [gr-qc]} \BibitemShut {NoStop}%
\bibitem [{\citenamefont {{Khan}}\ and\ \citenamefont {{Malik}}(2013)}]{4b}%
  \BibitemOpen
  \bibfield  {author} {\bibinfo {author} {\bibfnamefont {M.~S.}\ \bibnamefont
  {{Khan}}}\ and\ \bibinfo {author} {\bibfnamefont {M.~A.}\ \bibnamefont
  {{Malik}}},\ }\href {\doibase 10.1007/s10509-013-1547-6} {\bibfield
  {journal} {\bibinfo  {journal} {Astrophysics and Space Science}\ }\textbf
  {\bibinfo {volume} {348}},\ \bibinfo {pages} {211} (\bibinfo {year}
  {2013})}\BibitemShut {NoStop}%
\bibitem [{\citenamefont {{Wahid}}\ \emph {et~al.}(2011)\citenamefont
  {{Wahid}}, \citenamefont {{Ahmad}},\ and\ \citenamefont {{Nazir}}}]{Ahmad}%
  \BibitemOpen
  \bibfield  {author} {\bibinfo {author} {\bibfnamefont {A.}~\bibnamefont
  {{Wahid}}}, \bibinfo {author} {\bibfnamefont {F.}~\bibnamefont {{Ahmad}}}, \
  and\ \bibinfo {author} {\bibfnamefont {A.}~\bibnamefont {{Nazir}}},\ }\href
  {\doibase 10.1007/s10509-011-0634-9} {\bibfield  {journal} {\bibinfo
  {journal} {Astrophysics and Space Science}\ }\textbf {\bibinfo {volume}
  {333}},\ \bibinfo {pages} {241} (\bibinfo {year} {2011})}\BibitemShut
  {NoStop}%
\bibitem [{\citenamefont {{Ahmad}}\ \emph {et~al.}(2009)\citenamefont
  {{Ahmad}}, \citenamefont {{Wahid}}, \citenamefont {{Malik}},\ and\
  \citenamefont {{Masood}}}]{ce}%
  \BibitemOpen
  \bibfield  {author} {\bibinfo {author} {\bibfnamefont {F.}~\bibnamefont
  {{Ahmad}}}, \bibinfo {author} {\bibfnamefont {A.}~\bibnamefont {{Wahid}}},
  \bibinfo {author} {\bibfnamefont {M.~A.}\ \bibnamefont {{Malik}}}, \ and\
  \bibinfo {author} {\bibfnamefont {S.}~\bibnamefont {{Masood}}},\ }\href
  {\doibase 10.1142/S0218271809014364} {\bibfield  {journal} {\bibinfo
  {journal} {International Journal of Modern Physics D}\ }\textbf {\bibinfo
  {volume} {18}},\ \bibinfo {pages} {119} (\bibinfo {year} {2009})}\BibitemShut
  {NoStop}%
\bibitem [{\citenamefont {Hameeda}\ \emph {et~al.}(2018)\citenamefont
  {Hameeda}, \citenamefont {Upadhyay}, \citenamefont {Faizal}, \citenamefont
  {Ali},\ and\ \citenamefont {Pourhassan}}]{Hameeda}%
  \BibitemOpen
  \bibfield  {author} {\bibinfo {author} {\bibfnamefont {M.}~\bibnamefont
  {Hameeda}}, \bibinfo {author} {\bibfnamefont {S.}~\bibnamefont {Upadhyay}},
  \bibinfo {author} {\bibfnamefont {M.}~\bibnamefont {Faizal}}, \bibinfo
  {author} {\bibfnamefont {A.~F.}\ \bibnamefont {Ali}}, \ and\ \bibinfo
  {author} {\bibfnamefont {B.}~\bibnamefont {Pourhassan}},\ }\href {\doibase
  10.1016/j.dark.2018.02.001} {\bibfield  {journal} {\bibinfo  {journal} {Phys.
  Dark Univ.}\ }\textbf {\bibinfo {volume} {19}},\ \bibinfo {pages} {137}
  (\bibinfo {year} {2018})},\ \Eprint {http://arxiv.org/abs/1712.08591}
  {arXiv:1712.08591 [gr-qc]} \BibitemShut {NoStop}%
\end{thebibliography}%

\end{document}